\documentclass[aps,prx,longbibliography,preprint]{revtex4-1}

\usepackage[utf8]{inputenc}
\usepackage{amsmath, amssymb,graphicx}
\usepackage[cdot,mediumqspace,squaren]{SIunits}
\usepackage{hyperref}
\usepackage{xcolor}

\renewcommand{\vec}{\mathbf} 
\newcommand{\e}[1]{\text{e}^{#1}} 
\newcommand{\eps}{\varepsilon} 
\newcommand{\db}[1]{\vec{\overline{#1}}} 

\graphicspath{{./images/},{./},{../images/}}
\begin{document}

\title{Non-Hermitian Linear Electro-Optic Effect Through Interactions of Free and Bound Charges}
\author{Sylvain Lanneb\`{e}re\textsuperscript{1}}
\author{Nader Engheta\textsuperscript{2}}
\author{M\'{a}rio G. Silveirinha\textsuperscript{3}}
\email{To whom correspondence should be addressed:
mario.silveirinha@tecnico.ulisboa.pt}
 \affiliation{\textsuperscript{1}
Department of Electrical Engineering, University of Coimbra and
Instituto de Telecomunica\c{c}\~{o}es, 3030-290 Coimbra, Portugal}
\affiliation{\textsuperscript{2} Department of Electrical and Systems Engineering, University of Pennsylvania, Philadelphia, Pennsylvania 19104, USA}
\affiliation{\textsuperscript{3}University of Lisbon -- Instituto
Superior T\'ecnico and
Instituto de Telecomunica\c{c}\~{o}es, Department of Electrical Engineering, 1049-001
Lisboa, Portugal}

\begin{abstract}
In recent years, there has been growing interest in non-Hermitian phenomena in low-symmetry conductors, particularly optical gain driven by electro-optic effects. Conventional semiclassical treatments typically attribute these effects to nonlinear interactions associated with the anomalous velocity of Bloch electrons. Here, we present a phenomenological microscopic model that not only recovers these anomalous-velocity contributions, but also incorporates interband effects that become significant at higher frequencies. Our model captures a wide range of nonlinear interactions while remaining consistent with passivity and microscopic reversibility. Using this broader framework, we study the nonlinear interactions between free and bound electrons as an alternative mechanism for optical gain.
 We show that, under non-equilibrium conditions in low-symmetry conductors, the linearized electromagnetic response can exhibit both nonreciprocity and gain, even without anomalous velocity contributions. Finally, we analyze the stability of electrically biased systems and highlight potential applications such as optical isolators and traveling-wave amplifiers.
\end{abstract}

\maketitle

\section{Introduction}
Despite considerable research, photonic integrated circuits continue to face significant challenges due to losses at subwavelength scales, which restrict miniaturization below the diffraction limit \cite{chrostowski_silicon_2015, khurgin_scaling_2011, khurgin_how_2015}. This limitation has spurred the development of nanoscale light amplifiers to mitigate these losses \cite{chrostowski_silicon_2015, zhou_chip_2015, azzam_ten_2020}, yet there remains a critical need for enhanced electrically pumped devices, particularly in the terahertz domain. Moreover, the integration of efficient nonreciprocal components, such as isolators and circulators, continues to be challenging, despite their crucial role in the modular design of communication systems \cite{pozar_microwave_2011, levy_chip_2002, zhang_monolithic_2019, yang_nonreciprocal_2020, asadchy_tutorial_2020}.

In recent years, several magnetless mechanisms  to design circulators and isolators have emerged, including time modulation \cite{sounas_non-reciprocal_2017,williamson_integrated_2020,galiffi_broadband_2019,wang_nonreciprocity_2020,li_nonreciprocity_2022},
 nonlinear effects \cite{cotrufo_nonlinearity_2021,cotrufo_nonlinearity2_2021}, drifting electrons in high mobility materials \cite{borgnia_quasi-relativistic_2015,duppen_current-induced_2016,morgado_drift_2018,wenger_current-controlled_2018,bliokh_electric_2018,morgado_nonlocal_2020,morgado_active_2021,zhao_efficient_2021,dong_fizeau_2021}, 
 electrical or optical pumping \cite{buddhiraju_nonreciprocal_2020,lannebere_nonreciprocal_2022,rappoport_engineering_2023} and Weyl semimetals \cite{asadchy_subwavelength_2020,han_giant_2022,wu_tunable_2023,chistyakov_tunable_2023}. Despite their promise, these alternative approaches have not yet achieved the same robustness as Faraday isolators \cite{williamson_integrated_2020,cotrufo_nonlinearity_2021} and the quest for better nonreciprocal components continues.  

Additionally, there has been renewed interest in electro-optic effects, especially in achieving an optical gain response through electric field biasing \cite{lannebere_nonreciprocal_2022, rappoport_engineering_2023,shi_berry_2023}. Notably, combining nonlinearities with a static electric bias facilitates a transistor-like operation in bulk materials, characterized by both gain and nonreciprocity \cite{lannebere_nonreciprocal_2022}. Previous research has demonstrated that such distributed transistor responses can be induced in metallic, low-symmetry materials possessing significant Berry curvature dipoles, such as strained bilayer graphene and tellurium \cite{rappoport_engineering_2023, morgado_non-hermitian_2024}. These phenomena stem from the nonlinear interactions between free electrons and electromagnetic fields, based on the ``anomalous velocity" term associated with Bloch electrons.

Moreover, metallic or semimetallic systems without inversion symmetry are known to exhibit giant nonlinear optical responses \cite{wu_giant_2017,patankar_resonance-enhanced_2018,hu_giant_2023}. For example, Ref. \cite{wu_giant_2017} reports that transition metal monopnictides such as TaAs have a $\db{\chi}^{(2)}$ nonlinearity nearly an order of magnitude larger than that of GaAs, a widely used material in nonlinear optics. This enhancement originates from the itinerant nature of the free carriers, which enables significantly larger displacements under optical excitation than in dielectrics, where electrons are strongly bound to the atomic nuclei. A potential drawback, however, is that metallic-like systems may exhibit higher dissipation than dielectric materials.\\

Remarkably, the optical gain arising from the non-Hermitian electro-optic (NHEO) effect may exhibit chiral properties, typically governed by the polarization handedness. Specifically, the interactions between the medium and the wave are polarization-dependent: one particular polarization causes the material to act as a regular dissipative system, while an orthogonal polarization state induces gain \cite{lannebere_nonreciprocal_2022, rappoport_engineering_2023, morgado_non-hermitian_2024,lannebere_symmetry_2025}. Previous studies have shown that such indefinite-gain responses can be leveraged to develop innovative photonic devices with unique characteristics, including optical isolators with gain, chiral lasers, and polarization-dependent mirrors \cite{lannebere_nonreciprocal_2022, morgado_non-hermitian_2024, lannebere_chiral_2025, hakimi_chiral_2024}.

While the anomalous velocity contribution to electro-optic effects in low-symmetry conductors such as Weyl semimetals is already well understood, a phenomenological framework for capturing other contributions---particularly those involving interactions between free and bound electrons---remains underdeveloped.
In this work, we introduce a minimal Lorentz-type phenomenological model that describes such interactions. Heuristically, the proposed couplings are expected to capture nonlinear effects arising from interband processes in low-symmetry conductors.
Our analysis employs a semiclassical model, wherein free electrons are governed by standard transport and continuity equations, and bound charges follow Lorentzian dynamics. These dynamics are nonlinearly coupled through local fields, with the constraint that the interaction must satisfy to the principles of microscopic reversibility and passivity. We illustrate that under a static electric bias, the system’s linearized response is inherently non-Hermitian and nonreciprocal, with the type of response controlled by the symmetry of the material. Our model provides a foundation for interpreting nonlinear responses in low-symmetry conductors and for fitting experimental data as it becomes available.

The article is organized as follows. In section \ref{sec:microscopic_Berry}, we introduce a phenomenological nonlinear model for a material with free electrons possessing an anomalous velocity component. We demonstrate that the linearized optical response provided by the model agrees qualitatively with the results derived from semiclassical Boltzmann theory. In section \ref{sec:microscopic_model}, the model is extended to include both free and bound electrons. We analyze the possible forms of nonlinear coupling consistent with passivity and microscopic reversibility. We derive the general expression for the linearized permittivity response of the material. In section \ref{sec:spatial_symmetries}, we discuss various spatial symmetries compatible with nonreciprocal and gain responses, focusing on spatial symmetries of the \textit{2mm} point group and on the electro-optic response arising from interband interactions. We show that the NHEO effect in such systems originates a robust nonreciprocal and non-Hermitian electromagnetic response tailored by the electric field bias, even in the absence of an anomalous velocity term. In Section \ref{sec:wave_propagation}, we explore wave propagation and phenomena associated with the proposed models.  We show that materials belonging to the \textit{2mm} point group may enable the design of optical isolators and optical amplifiers.
Conclusions are drawn in Sec. \ref{sec:conclusion}.

\section{Phenomenological model for the NHEO effect with Berry curvature dipoles} \label{sec:microscopic_Berry}

To begin with, we introduce a phenomenological semiclassical model for the NHEO effect, considering only the contribution from free electrons to the response function. To this end, we consider an hypothetical system where the transport of electrons is described by a modified Drude-type model:
\begin{align}
\label{E:semiclassical1}
\frac{{d{\bf{p}}}}{{dt}} + \Gamma {\bf{p}} = q\left[ {{\bf{E}} + \frac{{\bf{p}}}{m} \times {\bf{B}}} \right].
\end{align}  
where $q=-e$ is the electron charge, $m$ is an effective mass, $\Gamma$ is a collision frequency, and $\bf{p}$ may be understood as a crystal quasi-momentum. It is supposed that due to the interactions with the ionic lattice the electron velocity and the crystal quasi-momentum are related as follows:
\begin{align}  
\label{E:semiclassical2}
{\bf{v}} = \frac{1}{m}\left[ {{\bf{p}} + \left( {\overline \zeta  \cdot {\bf{p}}} \right) \times {\bf{E}}} \right]
\end{align}
where $\overline \zeta$ is some system dependent tensor. The structure of the equations \eqref{E:semiclassical1} and  \eqref{E:semiclassical2} is reminiscent of the semiclassical equations for Bloch electrons in a crystal with a nontrivial Berry curvature \cite{xiao_berry_2010}. In particular, the nonlinear interaction of the electrons with the crystal lattice originates a contribution to the velocity which is proportional to the electric field $\bf{E}$. In the context of Bloch electrons, this term is known as the anomalous velocity contribution and depends on the geometry (Berry curvature) of the crystal bands \cite{xiao_berry_2010}. Equation \eqref{E:semiclassical2} is intended to effectively model the contribution of the Bloch electrons near the Fermi level, and thereby its structure differs slightly from that of the conventional anomalous velocity.

In our model, the transport equations \eqref{E:semiclassical1} and  \eqref{E:semiclassical2} are coupled to the Maxwell equations,
\begin{align}
\nabla  \times {\bf{E}} =  - {\mu _0}{\partial _t}{\bf{H}}, \quad \quad  \nabla  \times {\bf{H}} = {\varepsilon _{\text{b}}}{\partial _t}{\bf{E}}{\text{  + }}{\bf{j}}_{\text{c}}
\end{align}
through an electric current density,
\begin{align}
\label{E:currentdensity}
{\bf{j}}_{\text{c}} = {n_0}q{\bf{v}}
\end{align}
where $n_0$ is the electron density and $\varepsilon_{\text{b}}$ is the (non-dispersive) permittivity of the background due to the bound electrons.

\subsection{Passivity and conservation laws} 
\label{Sec:passivity}

Next, we show that the proposed model leads to a passive material response. To begin with, we note that from Eq.  \eqref{E:semiclassical2} it follows that ${\bf{E}} \cdot {\bf{v}} = {\bf{E}} \cdot \frac{{\bf{p}}}{m}$. Then, calculating the dot product of both members of \eqref{E:semiclassical1} with $\bf{p}$ it is found that 
${\bf{p}} \cdot \frac{{d{\bf{p}}}}{{dt}} + \Gamma {\bf{p}} \cdot {\bf{p}} = qm{\bf{E}} \cdot {\bf{v}}$. Thereby, the system satisfies:
\begin{align}
\label{Eq: free_electrons_conservationlaw}
\frac{d}{{dt}}\left[ {\frac{{{n_0}}}{{2m}}{\bf{p}} \cdot {\bf{p}}} \right] + \frac{{{n_0}\Gamma }}{m}{\bf{p}} \cdot {\bf{p}} = {\bf{E}} \cdot {\bf{j}}_{\text{c}}.
\end{align}
From the Poynting theorem \cite{kong_electromagnetic_1986}, it is well-known that $\nabla  \cdot {\bf{S}} + {\partial _t}{W_{{\text{EM}}}} =  - {\bf{E}} \cdot {\bf{j}}_{\text{c}}$, with ${\bf{S}} = {\bf{E}} \times {\bf{H}}$ the Poynting vector and ${W_{{\text{EM}}}} = \frac{1}{2}{\varepsilon _{\text{b}}}{\bf{E}} \cdot {\bf{E}} + \frac{1}{2}{\mu _{\text{0}}}{\bf{H}} \cdot {\bf{H}}$ the electromagnetic energy density. Combining this result with Eq. \eqref{Eq: free_electrons_conservationlaw}, it is found that the proposed phenomenological model adheres to the conservation law:
\begin{align}
\label{E:Poynting_freeelectrons}
\nabla  \cdot {\bf{S}} + {\partial _t}{W_{{\text{tot}}}} =  - Q,\quad \quad Q = \frac{{{n_0}\Gamma }}{m}{\bf{p}} \cdot {\bf{p}},\\
{W_{{\text{tot}}}} = \frac{1}{2}{\varepsilon _{\text{b}}}{\bf{E}} \cdot {\bf{E}} + \frac{1}{2}{\mu _{\text{0}}}{\bf{H}} \cdot {\bf{H}} + {n_0}\frac{1}{{2m}}{\bf{p}} \cdot {\bf{p}}.
\end{align}
The term ${n_0}\frac{1}{{2m}}{\bf{p}} \cdot {\bf{p}}$ is related to the kinetic energy density of the free electrons. 
Therefore, $W_{\text{tot}}$ represents the total energy density within the medium, and $Q > 0$ symbolizes the dissipated power per unit volume. This shows that our model describes a passive response. Particularly, in the absence of lattice collisions ($\Gamma \to 0^+$), the dynamics of the system are conservative, and the system's response remains time-reversal invariant in this limit. Note that the conservation law is independent of the tensor $\overline \zeta$ that governs the anomalous velocity contribution.

\subsection{Linearized dynamical response} \label{sec:linearized_response_free_electrons}

Let us now consider the scenario where the material is biased by a static electric field ${\bf{E}}_0$ created by some external generator. We write the total field as the sum the DC component and a weak dynamic field: ${\bf{E}} = {{\bf{E}}_0} + \left( {{{\bf{E}}_\omega }{e^{ - i\omega t}} + \text{ c.c.}} \right)$. As Eq. \eqref{E:semiclassical1} is linear, the corresponding quasi-momentum has a similar decomposition, ${\bf{p}} = {{\bf{p}}_0} + \left( {{{\bf{p}}_\omega }{e^{ - i\omega t}} + \text{c.c.}} \right)$ with  
\begin{align}
\label{E:responsep}
{{\bf{p}}_\omega } = \frac{q}{{\Gamma  - i\omega }}{{\bf{E}}_\omega },\quad \quad {{\bf{p}}_0} = \frac{q}{\Gamma }{{\bf{E}}_0}
\end{align}
For simplicity, we neglect the  magnetic field ($\bf{B}$) in the effective transport equation [Eq. \eqref{E:semiclassical1}]. It should be noted that the presence of collisions ($\Gamma\neq0$) is essential to ensure that the drift velocity (${{\bf{v}}_0} \approx {{\bf{p}}_0}/m$) remains finite.

Combining Eqs. \eqref{E:semiclassical2} and \eqref{E:currentdensity}, it is evident that the (linearized) dynamical current is given by ${{\bf{j}}_{{\text{c,}}\omega }} = \frac{{{n_0}q}}{m}\left[ {{{\bf{p}}_\omega } + \left( {\overline \zeta   \cdot {{\bf{p}}_\omega }} \right) \times {{\bf{E}}_0} + \left( {\overline \zeta   \cdot {{\bf{p}}_0}} \right) \times {{\bf{E}}_\omega }} \right]$. Then, from Eq. \eqref{E:responsep}, it follows that the linearized current is related to the dynamical field as ${{\bf{j}}_{{\text{c,}}\omega }} = \left( {{{\overline \sigma  }_{\text{D}}} + {{\overline \sigma  }_{{\text{EO}}}}} \right) \cdot {{\bf{E}}_\omega }$, where the ${{{\overline \sigma  }_{\text{D}}} + {{\overline \sigma  }_{{\text{EO}}}}}$ is the optical conductivity. It consists of a Drude-like response (${{\overline \sigma  }_{\text{D}}}$) and an electro-optic response that originates from the static-bias (${{\overline \sigma  }_{\text{EO}}}$):
\begin{align}
\label{E:sigmaEO}
{\overline \sigma  _{\text{D}}} = {\varepsilon _0}\omega _p^2\frac{1}{{\Gamma  - i\omega }}{{\bf{1}}_{3 \times 3}},\quad \quad \quad {\overline \sigma  _{{\text{EO}}}} = {\varepsilon _0}\omega _p^2\left[ {\frac{1}{\Gamma }\left( {\overline \zeta   \cdot {{\bf{E}}_0}} \right) \times {{\bf{1}}_{3 \times 3}} + \frac{{ - 1}}{{\Gamma  - i\omega }}{{\bf{E}}_0} \times \overline \zeta  } \right]
\end{align}
where ${\omega _p} = \sqrt {\frac{{{n_0}{q^2}}}{{{\varepsilon _0}m}}}$ is the plasma frequency of the electron gas. The electro-optic conductivity piece is proportional to the magnitude of the electric bias.

Interestingly, the term ${{\overline \sigma}_{\text{EO}}}$ shares the same structure as the electro-optic conductivity derived using the semiclassical Boltzmann formalism in \cite{morgado_non-hermitian_2024}. Indeed, the results from the phenomenological model align precisely with those calculated using Boltzmann's approach when the tensor $\overline \zeta$ satisfies:
\begin{align}
\overline \zeta   =  - \frac{{{e^3}}}{{{\varepsilon _0}\omega _p^2{\hbar ^2}}}{\overline {\bf{D}} ^T}.
\end{align}
Here, $\overline{\mathbf{D}}$ represents the Berry curvature dipole of the material. This similarity confirms that the phenomenological model describes precisely the electro-optic response predicted by the semiclassical Boltzmann theory. As extensively discussed in prior research \cite{morgado_non-hermitian_2024, rappoport_engineering_2023}, the electrically biased material may exhibit optical gain. Our phenomenological model elucidates how this property is consistent with the material’s global passivity [Eq. \eqref{E:Poynting_freeelectrons}] and highlights that the optical gain originates from the nonlinear interactions between DC and dynamic fields.

\subsection{Lagrangian formulation}

Importantly, as demonstrated in Appendix \ref{sec:Lagrangian}, linearizing any system described by a Lagrangian subjected to a time-independent bias---essentially, a system in equilibrium---inevitably results in a Hermitian linearized response. Consequently, the system of equations \eqref{E:semiclassical1}-\eqref{E:currentdensity} cannot be derived from a Lagrangian density, even when approaching the limit $\Gamma \to 0^+$. More broadly, we conclude that a nonlinear conservative system in equilibrium cannot exhibit gain upon linearization, thereby precluding its use for implementing the NHEO effect. Consistent with this finding, the analyses in \cite{morgado_non-hermitian_2024, rappoport_engineering_2023} demonstrate that the Berry curvature dipole can only be nonzero in metallic systems, specifically, systems where a static electric bias induces a drifting current, representing a nonequilibrium scenario.

Therefore, the NHEO effect can only occur in nonlinear systems out of equilibrium. As seen in Sec. \ref{Sec:passivity}, these systems can exhibit conservative-type dynamics (in the
sense that a Poynting theorem can be  written for the overall response), but their dynamics
do not derive from a Lagrangian. 

In this context, it is noteworthy that the anomalous velocity contribution in Eq. \eqref{E:semiclassical2} is represented by a tensor lacking transposition symmetry, $- \frac{1}{m}{\bf{E}} \times \overline \zeta$. In contrast, the term describing the effective mass is always a symmetric tensor (as exemplified here by $\frac{1}{m}{{\bf{1}}_{3 \times 3}}$, which corresponds to a scalar). Thus, the correction due to the anomalous velocity cannot be regarded as a field-dependent effective mass. This underscores the unique characteristics of the (nonlinear) term associated with the NHEO effect. 

\section{Model of free and bound charges nonlinearly coupled} \label{sec:microscopic_model}
Next, we explore alternative physical mechanisms that may potentially contribute to the NHEO effect, beyond the influence of the anomalous velocity. Specifically, we demonstrate that the NHEO effect can emerge from the interaction between nonlinearly coupled free and bound charges, even without the presence of an anomalous velocity term. The motivation for considering such interactions is to model interband effects of the electro-optic susceptibility response  $\db{\chi}_\text{EO}$.  Indeed, the  $\db{\chi}_\text{EO}$ computed from first principle approaches, such as via the Kubo formalism \cite{aversa_nonlinear_1995,jia_equivalence_2024}, can be decomposed as $\db{\chi}_\text{f-f}(\omega)+\db{\chi}_\text{f-b}(\omega)$ , where  $\db{\chi}_\text{f-f}(\omega)=\db{\sigma}_\text{EO}(\omega)/(-i\omega\eps_0)$ represents the piece of the electro-optic response associated with the Berry curvature dipole (semiclassical response, already discussed in Sect. \ref{sec:linearized_response_free_electrons}), whereas  $\db{\chi}_\text{f-b}(\omega)=\db{\sigma}_\text{EO}^\text{\,inter}(\omega)/(-i\omega\eps_0)$ represents the remaining part of the electro-optic response due to interband interactions. Note that for low frequencies, the semiclassical part of the response dominates because $\lim_{\omega \to 0^+}\db{\sigma}_\text{EO}^\text{\,inter}(\omega)=0$. However, for high frequencies, the additional interband contribution can become relevant, particularly near resonances. Our objective here is to introduce a phenomenological model that can describe  $\db{\chi}_\text{f-b}(\omega)$. 

Our microscopic model adheres to the principles of microscopic reversibility and passivity, even though the system dynamics is not described by a Lagrangian. As further discussed later, in principle, the parameters of the phenomenological model can be mapped to the (electro-optic) second-order susceptibilities obtained from band theory \cite{jia_equivalence_2024}. 
The magnitude of such interband effects in low-symmetry conductors \cite{wu_giant_2017,patankar_resonance-enhanced_2018,hu_giant_2023} may, in principle, be comparable to those of intraband NHEO effects (and, perhaps, even supplant them near resonances), making their study relevant for applications.

\subsection{Description of the model}\label{sec:description_model}
\begin{figure*}[!ht]
\centering
\includegraphics[width=0.45\linewidth]{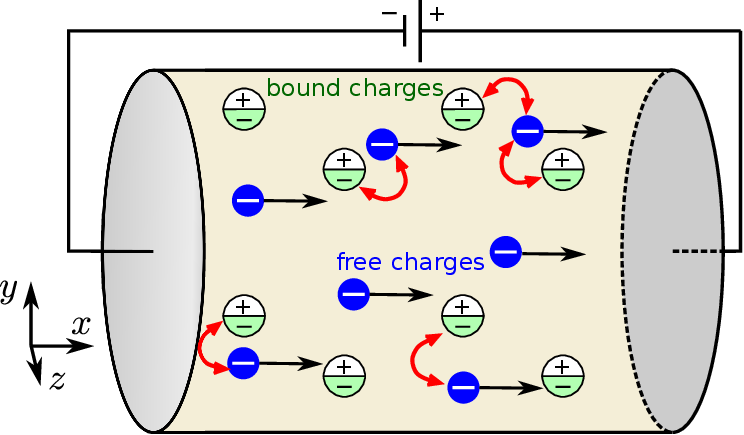}
       \caption{Schematic of a material formed by free carriers and bound charges that are nonlinearly coupled. The figure represents the situation where the system is biased by a static electric field  $\vec{E}_0$. For simplicity, only free carriers with a negative charge are represented.}
\label{fig:system_under_study}
\end{figure*}

We consider a low-symmetry conductor, schematically represented in Fig. \ref{fig:system_under_study}, that consists of free carriers and bound charges in interaction.
%

For simplicity, the free carriers are chosen to be  all of the same type with a mass $m$ and a charge $q$.
The bound charges are modelled through a standard anisotropic Lorentzian model. It is supposed that the two types of charges are coupled through local fields created by the conduction and polarization currents. Thus, the dynamics of the two types of carriers is described by:
\begin{gather}  \label{E:transport_eq_simple}
     \left( \partial_t + \Gamma  \right) \cdot \vec{p}  = q \left( \vec{E} + \db{C}_{12} \cdot \partial_t \vec{P}\right)  \\ \label{E:Lorentz}
m_\text{b} \left( \partial_{tt} + \gamma \partial_{t} + \sum_{j=x,y,z} \omega_{0j}^2 \hat{\vec{e}}_j\otimes \hat{\vec{e}}_j\right)  \vec{P} = q^2 n_\text{b} \left( \vec{E} + \db{C}_{21} \cdot \frac{q n_0}{m} \vec{p} \right)
\end{gather}
%
%
%
In the above, $\vec{P}$ is the polarization vector associated with the bound charges, $\omega_{0j}$ are the resonance frequencies (transverse optical phonon frequencies) along each axis, $\gamma$ is the phonon collision frequency, $m_\text{b}$ is the effective mass of the bound charges and $n_\text{b}$ their density. The notations $\partial_t \equiv \frac{\partial}{\partial t}$ and $\partial_{tt}\equiv \frac{\partial^2}{\partial t^2}$ are used for brevity. The vectors $\hat{\vec{e}}_x$, $\hat{\vec{e}}_y$, $\hat{\vec{e}}_z$ are the unit vectors along the Cartesian axes and $\otimes$ represents the tensor product of two vectors.
For simplicity, we neglect the effect of the magnetic component of the Lorentz force in Eq. \eqref{E:transport_eq_simple}, which is justified when the drift velocity of the free carriers satisfies $|\vec{v}_0|\ll c$. The quasi-momentum distribution $\vec{p}$ of the free charges is related to the drift velocity as in Eq. \eqref{E:semiclassical2}.

In our model, the free carriers and the bound charges are coupled  via two generic tensors, $\db{C}_{12} $ and $\db{C}_{21}$, with units of $\mathrm{Vm/A}$. In the absence of coupling, $\db{C}_{12}=\db{C}_{21}=0$, and of an anomalous velocity term, $\db{\zeta}=0$, equations \eqref{E:transport_eq_simple} and  \eqref{E:Lorentz} reduce to the standard Drude and Lorentz models. For completeness, in Appendix \ref{Sec:ApGeneralized} we present a slightly refined version of the proposed model that ensures that the conduction current satisfies the continuity equation. As discussed in Appendix \ref{Sec:ApGeneralized}, the corrections arising from such refinement are usually negligible.\\
The physical origin of the nonlinear coupling can be understood as follows: the current densities of bound charges $\partial_t \vec{P}$ and of free carriers $q n_0 \vec{v} \approx \frac{q n_0}{m} \vec{p}$ induced by an external field $\vec{E}$ generate secondary local electric fields given  by $\db{C}_{12} \cdot \partial_t \vec{P}$ and $\db{C}_{21} \cdot \frac{q n_0}{m} \vec{p}$, respectively. These secondary local fields give rise to forces (e.g., through the nonlinearity of the Lorentz force or nonclassical effects) that interact with the free and bound charges.
Physically, the coupling between the bound and free charges depends on the actual microscopic distribution of charges and fields. Thus, $\db{C}_{12} $ and $\db{C}_{21}$  may be functions of the different dynamic variables of the system.

It is relevant to note that previous works have considered nonlinear effects in the material response of metals \cite{scalora_second_2010,ciraci_hydrodynamic_2013}, including interactions between bound and free electrons. However, these typically involve couplings mediated by the Lorentz force, as in microscopic field theory. In contrast, our model introduces a more general class of nonlinear couplings, permissible within the framework of macroscopic electrodynamics, where the fields are effective and may include additional interaction terms arising from the peculiar physics of Bloch electrons.

\subsection{Allowed form of nonlinear coupling} \label{sec:allowed_form_coupling}

To make further progress, next we impose  restrictions on the allowed forms of the coupling  based on the passivity of the system.

The analysis of Sec. \ref{Sec:passivity} can be readily generalized to a material described by Eqs.  \eqref{E:transport_eq_simple}-\eqref{E:Lorentz}. To this end, first we calculate the inner product of both sides of Eq. \eqref{E:transport_eq_simple} with $\vec{p}$. Taking into account that ${\bf{E}} \cdot \frac{{\bf{p}}}{m} = {\bf{E}} \cdot {\bf{v}}$, it is found that:
\begin{align} 
\label{E:energyaux1}
\frac{\partial }{{\partial t}}\left[ {\frac{{{n_0}}}{{2m}}{\bf{p}} \cdot {\bf{p}}} \right] + \frac{{{n_0}\Gamma }}{m}{\bf{p}} \cdot {\bf{p}} = {\bf{E}} \cdot q{n_0}{\bf{v}} + \frac{{q{n_0}}}{m}{\bf{p}} \cdot {\overline {\bf{C}} _{12}} \cdot {\partial _t}{\bf{P}}.
\end{align}
Similarly, calculating the inner product of both sides of Eq. \eqref{E:Lorentz} with ${\partial _t}{\bf{P}}$, one obtains:
\begin{align} 
\label{E:energyaux2}
&\frac{\partial }{{\partial t}}\left[ {\frac{{{m_\text{b}}}}{{2{n_\text{b}}{q^2}}}{\partial _t}{\bf{P}} \cdot {\partial _t}{\bf{P}} + \frac{m_\text{b}}{{2{n_\text{b}}{q^2}}}\sum\limits_j {\omega _{0j}^2{\bf{P}} \cdot {{{\bf{\hat e}}}_j} \otimes {{{\bf{\hat e}}}_j}}  \cdot {\bf{P}}} \right] \\\nonumber &+ \gamma \frac{{{m_\text{b}}}}{{{n_\text{b}}{q^2}}}{\partial _t}{\bf{P}} \cdot {\partial _t}{\bf{P}} = {\bf{E}} \cdot {\partial _t}{\bf{P}} + {\partial _t}{\bf{P}} \cdot {\overline {\bf{C}} _{21}} \cdot \frac{{q{n_0}}}{m}{\bf{p}}
\end{align}
On the other hand, from the Poynting theorem we have that $\nabla  \cdot {\bf{S}} + {\partial _t}{W_{{\text{EM}}}} =  - {\bf{E}} \cdot {{\bf{j}}_{{\text{tot}}}}$, where $W_{\text{EM}}$ is defined as in Sec. \ref{Sec:passivity} and ${{\bf{j}}_{{\text{tot}}}}$ is the total current coupled to Maxwell's equations. It is now given by the sum of the conduction and polarization currents: ${{\bf{j}}_{{\text{tot}}}} = {n_0}q{\bf{v}} + {\partial _t}{\bf{P}}$. Using Eqs. \eqref{E:energyaux1}-\eqref{E:energyaux2}, it is readily found that:
\begin{align} 
\label{E:law_energy_conservation}
\nabla  \cdot {\bf{S}} + {\partial _t}{W_{{\text{tot}}}} =  - Q + \frac{{q{n_0}}}{m}\left[ {{\partial _t}{\bf{P}} \cdot {{\overline {\bf{C}} }_{21}} \cdot {\bf{p}} + {\bf{p}} \cdot {{\overline {\bf{C}} }_{12}} \cdot {\partial _t}{\bf{P}}} \right]
\end{align}
where we introduced
\begin{align} 
\label{E:Wtot}
{W_{{\text{tot}}}} &= {W_{{\text{EM}}}} + {\frac{{{n_0}}}{{2m}}{\bf{p}} \cdot {\bf{p}}} + \frac{{{m_\text{b}}}}{{2{n_\text{b}}{q^2}}}
\left(
{\partial _t}{\bf{P}} \cdot {\partial _t}{\bf{P}} + \sum\limits_j {\omega _{0j}^2{\bf{P}} \cdot {{{\bf{\hat e}}}_j} \otimes {{{\bf{\hat e}}}_j}}  \cdot {\bf{P}}
\right),\\
\label{E:Qtot}
Q &= \frac{{{n_0}\Gamma }}{m}{\bf{p}} \cdot {\bf{p}} + \gamma \frac{{{m_\text{b}}}}{{{n_\text{b}}{q^2}}}{\partial _t}{\bf{P}} \cdot {\partial _t}{\bf{P}}.
\end{align}
The terms of $W_{\text{tot}}$ represent the electromagnetic field energy density, the energy density of the free carriers, and the kinetic and potential energy density of the bound charges, respectively. 

The right-hand side of Eq. \eqref{E:law_energy_conservation} represents the power transferred from the material to the wave. For a passive system it must be strictly negative for any system state. This is possible only if the term in rectangular brackets vanishes. 
This can be enforced by imposing that
\begin{align} 
\db{C}_{12}^T=-\db{C}_{21}, \label{E:condition_conservative_coupling}
\end{align}
where the superscript $T$ represents the transpose operator.
When the above constraint is fulfilled, Eq. \eqref{E:law_energy_conservation} simplifies to the standard law of energy conservation \cite{pitaevskii_physical_2012} with a light and a matter part $\nabla  \cdot {\bf{S}} + {\partial _t}{W_{{\text{tot}}}} =  - Q$, where $Q>0$ represents the heat dissipated per unit of volume.
The condition \eqref{E:condition_conservative_coupling} will be assumed in the rest of the article. 

\subsection{Microscopic reversibility}\label{sec:reversibility}
The vast majority of physical systems obey the principle of microscopic reversibility, i.e. they are time-reversal (TR) invariant at the microscopic level \cite{sachs_physics_1987,asadchy_tutorial_2020}. Here, we impose that the coupling terms in Eqs. \eqref{E:transport_eq_simple} and \eqref{E:Lorentz} are compatible with such a principle and remain invariant under a TR symmetry transformation. 
The terms $\db{C}_{21} \cdot \vec{p} $ and $ -\db{C}_{21}^T  \cdot \partial_t \vec{P} $ must transform as ``forces'', and thus must be even under a TR symmetry transformation (similar to the other terms in the equations, with exception of the dissipative terms). Because $\vec{p}$ and $\partial_t \vec{P}$ are both odd  under a TR  symmetry transformation, it follows that $\db{C}_{21}$ must also be odd  under a TR, implying that $\db{C}_{21}$ cannot be a constant matrix. Thus, to respect microscopic reversibility, $\db{C}_{21}$ has to depend only on the state variables that are odd under a TR, namely: $\vec{v}$,  $\partial_t \vec{P}$, $\partial_t \vec{E}$ or $\vec{B}$.  
In particular, it follows that the coupling terms associated with the tensors  $\db{C}_{12}\cdot \partial_t \vec{P}$ and  $\db{C}_{21}\cdot \vec{p}$ represent nonlinear interactions between the bound and free charges. We will demonstrate later in Sec. \ref{sec:spatial_symmetries} that the
linearization of these nonlinear terms at the bias point may lead to a nonreciprocal and non-Hermitian electromagnetic response, even in the absence of a Berry curvature dipole $\left(\db{\zeta}=0\right)$.

Since we are only interested in the linearized electromagnetic response of the system, we can restrict the expansion of $\db{C}_{21}$ to \textit{linear} combinations of the variables odd under a TR symmetry ($\vec{v}$,  $\partial_t \vec{P}$, $\partial_t \vec{E}$ and $\vec{B}$).
As previously noted, the interactions between the free and bound charges are expected to depend on the dynamical state of the system, i.e., on the displacements of the two charge distributions from the equilibrium point. Thereby, it is reasonable to assume that $\db{C}_{21}$ is a linear function of $\vec{v}$ and  $\partial_t \vec{P}$ only. 
In these conditions, the coupling matrix is of the form: 
\begin{align} 
\db{C}_{21}&=\sum_{ijl} \left( a_{ijl}  \hat{\vec{e}}_l \cdot \vec{v}  + b_{ijl} \hat{\vec{e}}_l  \cdot \partial_t \vec{P} \right) \hat{\vec{e}}_i \otimes \hat{\vec{e}}_j  \label{E:C21}  
\end{align}
%
%
where the dummy indices $i$, $j$ and $l$ run over $x,y,z$ and the coupling coefficients $a_{ijl}$ and $b_{ijl}$ are real-valued coefficients (54 in total) with units of $\mathrm{Vs/A}$ and $\mathrm{m^3V/A^2}$, respectively. Note that in the lossless limit ($\gamma \to 0^+$ and $\Gamma \to 0^+$) the full dynamical equations become TR-symmetric. 
The parameters in this phenomenological model are understood as effective quantities, analogous to the effective mass in band theory. 

\subsection{Linearization} 
\label{sec:linearized_permittivity}

Next, we linearize the nonlinearly coupled differential equations \eqref{E:transport_eq_simple} to \eqref{E:Lorentz} with the coupling matrix $\db{C}_{21}$ given by Eq. \eqref{E:C21}. The nonlinear response is determined both by $\db{C}_{21}$ and by the ``anomalous'' velocity term described by the tensor $\db{\zeta}$. 
 For weak nonlinearities, the two contributions to the linear electro-optic effect are independent and are combined additively. The effect of the anomalous velocity term was already characterized in Sec. \ref{sec:microscopic_Berry} [see Eq. \eqref{E:sigmaEO}]. Thus, to determine the contribution from the interactions of free and bound electrons, we can set $\db{\zeta}=0$ and use $\vec{p} = m \vec{v}$ in Eqs. \eqref{E:transport_eq_simple}-\eqref{E:Lorentz}. This leads to the following system of differential equations:
\begin{subequations}\label{E:nonlinear_eq_explicit_coupling}
\begin{align}  \label{E:eq_transport_2}
    m  \left( \partial_t + \Gamma  \right) \cdot \vec{v} & =  q \left( \vec{E} - \sum_{ijl} \left(a_{ijl} \hat{\vec{e}}_l \cdot \vec{v} + b_{ijl}\hat{\vec{e}}_l  \cdot \partial_t \vec{P}  \right) \hat{\vec{e}}_j \otimes \hat{\vec{e}}_i  \cdot \partial_t \vec{P}\right)   \\ 
m_\text{b} \left( \partial_{tt}+ \gamma \partial_{t} + \sum_j \omega_{0j}^2 \hat{\vec{e}}_j\otimes \hat{\vec{e}}_j \right)  \vec{P} &= q^2 n_\text{b} \left( \vec{E} + \sum_{ijl} \left(a_{ijl} \hat{\vec{e}}_l \cdot \vec{v}+ b_{ijl}\hat{\vec{e}}_l  \cdot \partial_t \vec{P}  \right) \hat{\vec{e}}_i \otimes \hat{\vec{e}}_j \cdot q n_0 \vec{v} \right)\label{E:eq_Lorentz_2}
\end{align}
 \end{subequations}
Let us first ignore the weak nonlinear terms associated with the coefficients $a_{ijl}$ and $b_{ijl}$.Then, the velocity and polarization vector response to a time-harmonic field (${\bf{E}} = {{\bf{E}}_\omega }{e^{ - i\omega t}}$) is determined by:
\begin{subequations}\label{E:linear_bound_free}
\begin{align}  
&{\bf{v}}_\omega ^0 = \frac{q}{m}\frac{1}{{\Gamma  - i\omega }}{{\bf{E}}_\omega },   \\
&{\bf{P}}_\omega ^0 = \eps_0 \sum\limits_j {\frac{\omega_\text{b}^2}{{\omega _{0j}^2 - \omega \left( {\omega  + i\gamma } \right)}}{{{\bf{\hat e}}}_j} \otimes {{{\bf{\hat e}}}_j}}  \cdot {{\bf{E}}_\omega }
\end{align}
 \end{subequations}
 where $\omega_\text{b}^2=\frac{q^2 n_\text{b}}{\eps_0 m_\text{b}}$.
In particular, the response to a static bias (${\bf{E}} = {{\bf{E}}_0 }$) is given by:
\begin{subequations}
\begin{align}  
     \vec{v}_0 & =  \frac{q}{m   \Gamma } \vec{E}_0,  \\  
\vec{P}_0 &= \eps_0\sum_j  \frac{ \omega_\text{b}^2 }{  \omega_{0j}^2 }\hat{\vec{e}}_j\otimes \hat{\vec{e}}_j\cdot   \vec{E}_0 .     
\end{align}
\end{subequations}
The static electric bias $\vec{E}_0$ induces a drift current characterized by the velocity $\vec{v}_0$, as well as a macroscopic polarization vector.

To perform the linearization of equations \eqref{E:nonlinear_eq_explicit_coupling}, it is assumed that a generic dynamic variable of the system $\vec{K}$---where $\vec{K}$ represents $\vec{v},\,\vec{P}$ or $\vec{E}$---can be decomposed into $\vec{K}=\vec{K}_0 + {\vec{K}_{\omega} } {e^{ - i\omega t}}$ where $\vec{K}_0$ is a time-independent vector induced by the static electric bias and $\vec{K}_{\omega} {e^{ - i\omega t}}$ is a time harmonic vector associated with the dynamical field. Similar to \cite{lannebere_nonreciprocal_2022}, it is assumed that the time-harmonic signal has a small amplitude with respect to the static bias such that $|\vec{K}_{\omega}| \ll |\vec{K}_0|$.

Substituting $\vec{K}=\vec{K}_0 + {\vec{K}_{\omega} } {e^{ - i\omega t}}$ in Eqs. \eqref{E:nonlinear_eq_explicit_coupling} and retaining only the terms linear in the dynamical fields one obtains:
 \begin{subequations}\label{E:diff_eq_linearized_0}
\begin{align}  \label{E:transport_first_order}
     & m  \left( -i \omega + \Gamma  \right) {\vec{v}_{\omega}}  =  q \left( {\vec{E}_{\omega}} +i \omega \sum_{jlr} a_{jlr} v_{0r} \hat{\vec{e}}_l \otimes \hat{\vec{e}}_j  \cdot  {\vec{P}_{\omega}}\right)  \\ 
    \sum_j & \left( \omega_{0j}^2-\omega^2 - i \gamma \omega \right) \hat{\vec{e}}_j\otimes \hat{\vec{e}}_j\cdot   {\vec{P}_{\omega}} =  \\ \nonumber   
& \hspace{3cm} \eps_0  \omega_\text{b}^2 \left( {\vec{E}_{\omega}} + q n_0 \sum_{jlr} v_{0r}    \hat{\vec{e}}_j \otimes \hat{\vec{e}}_l \cdot  \left[ \left(  a_{jlr} + a_{jrl} \right)   {\vec{v}_{\omega}} -i \omega b_{jrl}       {\vec{P}_{\omega}}       \right]\right)   
\end{align}
\end{subequations}
where $v_{0r}=\vec{v}_0 \cdot \hat{\vec{e}}_r$.

The linearized permittivity tensor $\db{\eps}$ is defined through the relation between the dynamic electric field ${\vec{E}}_{\omega}$ and the total dynamic polarization   ${\vec{P}}_{\text{tot},\omega}=\frac{n_0q}{- i \omega}{\vec{v}_{\omega}}  + {\vec{P}_{\omega}}$: 
\begin{align} 
\label{E:def_linearized_permittivity}
{\vec{P}}_{\text{tot},\omega}=\eps_0 \left( \db{\eps}-\vec{1}_{3\times3} \right) \cdot {\vec{E}}_{\omega}.
\end{align}
For a weak nonlinearity, ${\vec{v}_{\omega}}$ and ${\vec{P}_{\omega}}$ can be calculated using the replacements ${{\bf{v}}_\omega } \to {\bf{v}}_\omega ^0$ and
${{\bf{P}}_\omega } \to {\bf{P}}_\omega ^0$ in the right-hand sides of Eqs. \eqref{E:diff_eq_linearized_0}, with ${\bf{v}}_\omega ^0$ and ${\bf{P}}_\omega ^0$ given by Eqs. \eqref{E:linear_bound_free}. A straightforward analysis shows that the permittivity can be written as:
\begin{align}
\label{E:def_linearized_permittivity_general_case}
\db{\varepsilon} = \db{\varepsilon}_{\text{L}} + \db{\chi}_{\text{f-b}} + \db{\chi}_{\text{f-f}},   
\end{align}
where $\db{\eps}_{\text{L}}$ determines the material response without the bias,  
\begin{subequations}
\begin{align}  
\db{\eps}_{\text{L}} &= \sum_j \eps_{jj}^0 \hat{\vec{e}}_j \otimes\hat{\vec{e}}_j \\
\eps_{jj}^0(\omega) & =1 -   \frac{\omega_p^2}{\omega \left( \omega +i \Gamma  \right)} +  \frac{  \omega_\text{b}^2 }{\omega_{0j}^2 -\omega^2 -i\omega \gamma}  \label{E:eps_jj_0}
\end{align} 
\end{subequations}
and $\db{\chi}_{\text{f-b}}$ and $\db{\chi}_{\text{f-f}}$ represent the electro-optic response due to the interaction of bound and free electrons and due to the ``anomalous velocity", respectively:
\begin{subequations}
\begin{align}
\label{E:chibf}
\db{\chi}_{\text{f-b}} &= - i  \sum_{jlr} \left[  g_{lr}  a_{ljr}     -  g_{jr}  \left(  a_{jlr} + a_{jrl} \right)      -  f_{jlr}    b_{jrl}    \right]\hat{\vec{e}}_j \otimes\hat{\vec{e}}_l  \\
\db{\chi}_{\text{f-f}} &=
\frac{{{{\db{\sigma}  }_{{\text{EO}}}}}}{{ - i\omega {\varepsilon _0}}}.
\end{align}
\end{subequations}
The term due to the anomalous velocity was added by hand to the final result, because as noted previously the different nonlinear contributions combine additively. In the above, the $\omega$-dependent functions $g_{jr} $ and $f_{jlr}$ are defined by 
\begin{subequations}\label{E:functions_linearized_permittivity}
\begin{align} 
f_{jlr}(\omega)&= -  \eps_0 q n_0 v_{0r} \omega    \frac{  \omega_\text{b}^2 }{\omega_{0j}^2 -\omega\left(\omega +i \gamma\right) }  \frac{  \omega_\text{b}^2 }{\omega_{0l}^2 -\omega\left(\omega +i \gamma\right) }    \label{E:f_jlr},   \\ 
 g_{jr}(\omega) &= \eps_0v_{0r}    \frac{ \omega_p^2 \omega_\text{b}^2 }{\left( \omega +i \Gamma  \right)\left[ \omega_{0j}^2 -\omega\left(\omega +i \gamma\right) \right]}.  \label{E:g_jr} 
\end{align}
\end{subequations}
The linearized permittivity  \eqref{E:def_linearized_permittivity_general_case} is a dispersive function that in the absence of coupling ($a_{ijk}=b_{ijk}=0$ for all indices and $\db{\zeta}=0$) reduces to a diagonal matrix. As seen, the electro-optic effect may induce non-diagonal terms in the permittivity tensor. The term $\db{\chi}_{\text{f-b}}$ represents the part of the electro-optic response due to the interband effects that are not captured by the usual semiclassical calculation rooted on the Berry dipole ($\db{\chi}_{\text{f-f}}$). Note that $\db{\chi}_{\text{f-f}}$ has a pole in the static limit, whereas $\db{\chi}_{\text{f-b}}$ is finite in the same limit, confirming that the interband contribution is mostly relevant at high frequencies.

Remarkably, as discussed in the next section, for low-symmetry materials such non-diagonal terms may result in a permittivity tensor lacking transpose symmetry and may give rise to nonreciprocity and optical gain.
In the rest of the article, we focus our analysis in the free and bound electrons (interband) interactions, and thereby we will ignore possible anomalous velocity contributions $\left(\db{\zeta}=0\right)$.

\section{Spatial symmetries} 
\label{sec:spatial_symmetries}

 Next, we discuss the spatial symmetries compatible with nontrivial nonlinear couplings and derive explicitly the linearized permittivity for materials belonging to the \textit{2mm} point group.\\ 

\subsection{Constraints due to spatial symmetries} 

Next, we study how the invariance of the material under rotations, inversions, or their combinations, constrains the elements of the tensor $\db{C}_{21}$.

The tensor $\db{C}_{21}$ links ``acceleration'' and ``velocity" distributions. Both acceleration and velocity transform identically under rotations and reflections: 
$\vec{A}(\vec{r}) \to \vec{A}'(\vec{r'}) = \mathcal{R}\cdot\vec{A}(\mathcal{R}^T\cdot\vec{r})$
with $\vec{A}$ standing for the relevant field. Here $\mathcal{R}$ is an element of the orthogonal group $O(3)$, so that ${{\mathcal R}^T} \cdot {\mathcal R} = {{\bf{1}}_{3 \times 3}}$. Thus, if $\mathcal{R}$ is a symmetry of the material, the coupling matrix $\db{C}_{21}\left(   \vec{v},\partial_t \vec{P}\right)$ is constrained to satisfy:
\begin{align} 
\db{C}_{21}\left( \mathcal{R}\cdot \vec{v},\mathcal{R}\cdot \partial_t \vec{P} \right)&=\mathcal{R}\cdot\db{C}_{21}\left(   \vec{v},\partial_t \vec{P}\right) \cdot \mathcal{R}^T
\end{align}
This invariance under specific spatial symmetries restricts the form of the coupling tensor, potentially reducing the number of nonzero coupling coefficients significantly.

For example, for a system invariant under inversion symmetry (corresponding to $\mathcal{R}=-\vec{1}_{3\times 3}$) the above equation combined with Eq. \eqref{E:C21} implies that 
 $\db{C}_{21} = 0$. This means that a non-centrosymmetric structure is required to implement the NHEO effect.  

Among the non-centrosymmetric systems that possess an axis of rotation, it can be checked that rotations by an angle of $\pi$, $2\pi/3$, $\pi/2$, and $\pi/3$ are all compatible with the existence of a nontrivial coupling tensor $\db{C}_{21}$. 
For example, we checked that a nontrivial $\db{C}_{21}$ is   compatible with the symmetries of  the \textit{2}, \textit{222}, \textit{2mm} and \textit{32} crystallographic point groups (Hermann–Mauguin notation).

\subsection{Linearized permittivity for the \textit{2mm} point group} \label{sec:linearized_permittivity_2mm}
In the rest of the paper, we focus on material belonging to the \textit{2mm} point group. Materials in this crystallographic point group have two mirror planes and are invariant under a 2-fold rotation symmetry about some specific axis. We note in passing that a MOSFET transistor exhibits the same type of symmetries \cite{lannebere_nonreciprocal_2022}.
 
We   assume, without loss of generality, that the axis of rotation of the structure is parallel to the $z$ direction. The nontrivial elements of the \textit{2mm} point group consist of a rotation by an angle of $\pi$ ($\mathcal{R}=\text{diag}\left\{ -1,-1,1\right\}$) and two mirror symmetries with respect to planes containing the rotation axis (here taken as $\mathcal{R}=\text{diag}\left\{ -1,1,1\right\}$ and $\mathcal{R}=\text{diag}\left\{ 1,-1,1\right\}$). When these three symmetries are individually imposed to our model, it can be shown that most coupling coefficients in Eq.  \eqref{E:C21} must vanish. The most general form of the coupling tensor is:
 \begin{align}\label{E:C21_2mm_group} \db{C}_{21}^{2mm}=
\begin{pmatrix}
 a_{xxz} v_z + b_{xxz} \partial_t P_z    & 0 & a_{xzx} v_x+ b_{xzx}\partial_t P_x   \\
 0 & a_{yyz} v_z+ b_{yyz} \partial_t P_z   & a_{yzy} v_y+ b_{yzy} \partial_t P_y   \\
 a_{zxx} v_x+ b_{zxx} \partial_t P_x   & a_{zyy} v_y+ b_{zyy} \partial_t P_y   & a_{zzz} v_z+ b_{zzz} \partial_t P_z   \\
\end{pmatrix}  
 \end{align}
Thus, only 14 coupling coefficients out of the 54 possible are allowed to be different from zero.

Next, we calculate the linearized permittivity \eqref{E:def_linearized_permittivity_general_case} for a material that exhibits $2mm$ point group symmetry. Importantly, two cases can be distinguished depending on whether the direction of the drift velocity is parallel or orthogonal to the axis of rotation of the structure. For a drift velocity parallel to the axis of rotation ($\vec{v}_0=v_{0z}\hat{\vec{e}}_z$), it can be shown that the linearized permittivity \eqref{E:def_linearized_permittivity_general_case} is always a diagonal tensor and therefore describes a reciprocal response. This case will not be further considered in this paper. 

Instead, we focus on the situation where the drift velocity is orthogonal to the rotation axis and assume, without loss of generality, that it is along the $x$ direction: $\vec{v}_0=v_{0x}\hat{\vec{e}}_x$.
It is interesting to examine further the set of linearized differential equations \eqref{E:diff_eq_linearized_0} for this example. It reduces to:
 \begin{subequations}\label{E:diff_eq_linearized}
\begin{align} 
 m  \left( -i \omega  + \Gamma \right){\vec{v}_{\omega}}    & = q \left[  {\vec{E}_{\omega}}  + i \omega v_{0x}  (a_{zxx} \hat{\vec{e}}_x \otimes \hat{\vec{e}}_z + a_{xzx}  \hat{\vec{e}}_z \otimes \hat{\vec{e}}_x)  \cdot  {\vec{P}_{\omega}} \right]     \\
\sum_j  \left( \omega_{0j}^2-\omega^2 - i \gamma \omega \right) \hat{\vec{e}}_j\otimes \hat{\vec{e}}_j\cdot   {\vec{P}_{\omega}} &=\eps_0  \omega_\text{b}^2 \left\{ {\vec{E}_{\omega}}   +    q n_0  v_{0x} \left[ ( 2 a_{zxx} \hat{\vec{e}}_z \otimes \hat{\vec{e}}_x + \left( a_{xzx} + a_{xxz} \right) \hat{\vec{e}}_x \otimes \hat{\vec{e}}_z) \cdot    {\vec{v}_{\omega}}  \right.\right. \nonumber \\  & \hspace{3.2cm} \left.\left. -i \omega \left( b_{xxz} \hat{\vec{e}}_x \otimes \hat{\vec{e}}_z+ b_{zxx} \hat{\vec{e}}_z \otimes \hat{\vec{e}}_x \right) \cdot  {\vec{P}_{\omega}} \right] \right\}
\end{align}
\end{subequations}
This set of equations explicitly shows the form of linearized local field induced by the coupling between the free and bound electrons. As seen, the induced local fields involve the interaction between the $x$ component of $\vec{v}_\omega$ or $-i \omega\vec{P}_\omega$ with the $z$ component of $\vec{v}_\omega$ or $-i \omega \vec{P}_\omega$, or vice-versa, similar to what happens in standard MOSFET transistors \cite{sze_physics_2021,lannebere_nonreciprocal_2022}.

Furthermore, it is clear from the above equation that for a  drift velocity orthogonal to the material rotation axis, only 5 coefficients out of the 14 coefficients appearing in Eq. \eqref{E:C21_2mm_group} play a role in the linearized response. The other coefficients give rise to higher order responses and disappear upon linearization. 
%
 %
In fact, from \eqref{E:def_linearized_permittivity_general_case} and \eqref{E:chibf}, one readily finds that for materials belonging to the \textit{2mm} point group with a drift orthogonal to the rotation axis, the linearized permittivity tensor is given by:
\begin{align} \label{E:linearized_permittivity}
  \db{\eps} &= \begin{pmatrix}\eps_{xx}^0 && 0&& i \left(f_1 b_{xxz} -g_{zx} a_{zxx}+ g_{xx}\left[ a_{xzx}+ a_{xxz} \right]\right)\\ 0 && \eps_{yy}^0  && 0 \\ i \left( f_1b_{zxx} +  2 g_{zx} a_{zxx}-g_{xx}a_{xzx}\right)&& 0 && \eps_{zz}^0 \end{pmatrix}         
\end{align}
where $f_1(\omega)=f_{xzx} = f_{zxx}$ and the remaining functions are given by Eq. \eqref{E:functions_linearized_permittivity}. 

In the presence of nonlinear coupling between the free and bound electrons, when some of the coefficients $a_{ijk}$ and $b_{ijk}$ are different from zero, the permittivity tensor is non-diagonal. Additionally, due to collisions ($\Gamma > 0$ and $\gamma > 0$), the functions $\eps_{ii}^0$, $f_1$ and $g_{ij}$ (with $i,j=x,y,z$) are all complex-valued.
Remarkably, all types of coupling result in a permittivity tensor that is asymmetric, leading to a nonreciprocal ($\db{\eps}\neq \db{\eps}^T$) electromagnetic response. In particular, the linearized electro-optic response is not TR symmetric ($\db{\chi}_{\text{f-b}}\neq \db{\chi}_{\text{f-b}}^\ast$); note that the tensor $\db{\chi}_{\text{f-b}}$ is formed by the out-of-diagonal elements of $\db{\eps}$. This behavior is expected from the presence of a drift current, which is inherently odd under TR symmetry. 
The broken TR symmetry distinguishes our system from the idealized model described in Ref. \cite{lannebere_nonreciprocal_2022}, aligning it more closely with the behavior observed in actual MOSFET transistors, which also support a DC current \cite{fernandes_exceptional_2024}.

Additionally, 
all couplings
result in a non-Hermitian electro-optic response ($\db{\chi}_{\text{f-b}}\neq \db{\chi}_{\text{f-b}}^\dagger$), Furthermore, it can be checked that the tensor $\db{\chi}_\text{NH} = \frac{1}{2i} \left( \db{\chi}_\text{f-b} - \db{\chi}_\text{f-b}^\dag  \right)$ is an indefinite matrix. This indicates that the electro-optic interaction may result in either dissipation or in gain \cite{lannebere_chiral_2025}. This idea will be further developed in the next section.

Therefore, the nonlinear interactions between the bound and free electrons provide an alternative mechanism to implement the NHEO effect, not directly relying on the Berry curvature dipole. It is interesting to note that the frequency dependence of the tensor $\db{\chi}_{\text{f-b}}$ is qualitatively similar to that of the $\db{\chi}^{(2)}$ tensor derived from band theory \cite{aversa_nonlinear_1995,shen_principles_2003,boyd_nonlinear_2008} (Sec. 2.2 and 3.2, respectively). Thus, in principle, it is possible to link the coefficients $a_{ijk}$ and $b_{ijk}$ with the electronic band-structure of the material. Even though no data are currently available to enable such a mapping for materials of primary interest, we hope that the present model can serve as a foundation for future theoretical or experimental investigations.


It is relevant to note that the coupling associated with the coefficient $a_{xzx}$, gives rise to a gyrotropic response, reminiscent of that of a magnetized plasma \cite{pitaevskii_physical_2012,altman_reciprocity_2011}. However, different from passive gyrotropic media, for a sufficiently strong electric bias this coupling can originate chiral gain, analogous to the case of tellurium discussed in Ref. \cite{morgado_non-hermitian_2024}.

In summary, in this section it was demonstrated that a metallic system with the symmetry of the \textit{2mm} point group, with the static electric bias orthogonal to the rotation axis of the structure, can be used to implement the NHEO effect and achieve nonreciprocity and gain, exclusively through the nonlinearity arising from interband interactions not captured by the anomalous velocity. Different from other platforms based on drifting electrons \cite{borgnia_quasi-relativistic_2015,duppen_current-induced_2016,morgado_drift_2018,wenger_current-controlled_2018,bliokh_electric_2018,morgado_nonlocal_2020,morgado_active_2021,zhao_efficient_2021,dong_fizeau_2021}, our solution does not rely on Doppler shifts and on waves with extremely short wavelengths (plasmons). Furthermore, in our system the drift velocity does not need to be a significant fraction of the light velocity. 

We expect that our model can be used to describe the physics of electro-optic effects due to interband interactions (not captured by the semiclassical Boltzmann approach) in low-symmetry metals, in material platforms such as Weyl semimetals \cite{guo_light_2023}, transition metal dichalcogenides (TMDs), chiral perovskites \cite{long_chiral-perovskite_2020} or polar/ferroelectrics metals (materials that are both conducting and polarizable) \cite{zhou_review_2020,hickox-young_polar_2023,bhowal_polar_2023}, and engineered semiconductor superlattices \cite{rosencher_quantum_1996}. A particularly interesting candidate material is WTe$_2$, a TMD that is simultaneously a polar metal and a Weyl semimetal with the symmetry of the \textit{2mm} point group.

\section{Wave propagation} \label{sec:wave_propagation}
To illustrate the properties of our system, next we study the wave propagation in a material described by the linearized permittivity \eqref{E:linearized_permittivity}.  For simplicity, we assume an isotropic Lorentz model such that $\omega_{0j}=\omega_{0}$ for all $j$. Also we focus  on a specific type of nonreciprocal and non-Hermitian response and assume that $a_{zxx}$ is the only nonzero coupling coefficient in Eq. \eqref{E:C21}. Thus, the coupling matrix is $\db{C}_{21}=a_{zxx} v_x \hat{\vec{e}}_z\otimes \hat{\vec{e}}_x$, so that it couples the $x$ and $z$ components of the velocity and polarization current. From  Eq. \eqref{E:linearized_permittivity}, the linearized permittivity satisfies
\begin{align} \label{E:epsilon_example}
  \db{\eps} &= \begin{pmatrix}\eps_d && 0 && - i \eps_c \\ 0 && \eps_d && 0 \\ i 2 \eps_c && 0 && \eps_d \end{pmatrix}         
\end{align}
where $\eps_d=\eps_{jj}^0(\omega)$  and $\eps_c=g_{zx}(\omega) a_{zxx}$, with $\eps_{jj}^0$ and $g_{zx}$ given by Eq.\eqref{E:functions_linearized_permittivity}. 
In the rest of this section, we are interested only in weak
signals described by the linearized response, and thus, to
keep the notations short, we drop the subscript  $\omega$  (${
\vec{E}_\omega} \to \vec{E}$), and denote the dynamic electric field simply
by $\vec{E}$.

\subsection{Band structure and stability}
\label{Sec:stability}
To begin with, we characterize the photonic band diagrams in the absence of coupling ($\eps_c=0$). In this case, the permittivity tensor is isotropic $\db{\eps}=\eps_d \vec{1}_{3\times 3}$  and the plane wave solutions of the homogeneous wave equation $\nabla \times \nabla \times  \vec{E}
=  \frac{\omega^2}{c^2} \eps_d \vec{E}$ can be separated into a longitudinal mode and a doubly degenerated transverse electromagnetic (TEM) mode.
\begin{figure*}[!ht]
\centering
\includegraphics[width=\linewidth]{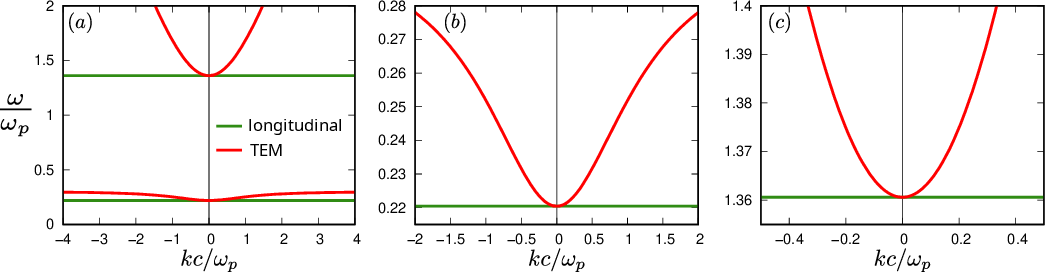}
       \caption{(a) Band diagram of the unbiased material showing the frequency as a function of the normalized wavenumber for $\omega_\text{b}=0.9 \omega_p$ and $\omega_0=0.3 \omega_p$. (b) Zoom in near the lower edge of the low frequency band. (c) Zoom in near the lower edge of the high frequency band. The horizontal green solid lines represent the longitudinal mode and the red lines represent the doubly degenerated TEM modes.}
\label{fig:band_diagram_no_coupling}
\end{figure*}
The photonic dispersion of these modes is represented in Fig. \ref{fig:band_diagram_no_coupling} in the weak dissipation limit ($\gamma$ and $\Gamma \to 0^+$). The dispersion consists of two distinct branches separated by a band gap. Each branch is formed by a doubly degenerate TEM band characterized by a wavevector $k = \frac{\omega}{c} \sqrt{\eps_d }$ and by a dispersionless longitudinal band. The longitudinal mode frequencies, $\omega_{lj}$, with $j=1,2$ are the solutions of $\eps_d(\omega_{lj})=0$. It is assumed that $0<\omega_{l1}<\omega_{l2}$.  The low-frequency band response is mainly determined by the bound charges, and is defined by $\omega_{l1}<\omega<\omega_0$, where $\omega_0$ is the polaritonic resonance. On the other hand, the high-frequency band, is mainly associated with the free charges, and is defined by $\omega>\omega_{l2}$.

\begin{figure*}[!ht]
\centering
\includegraphics[width=0.8\linewidth]{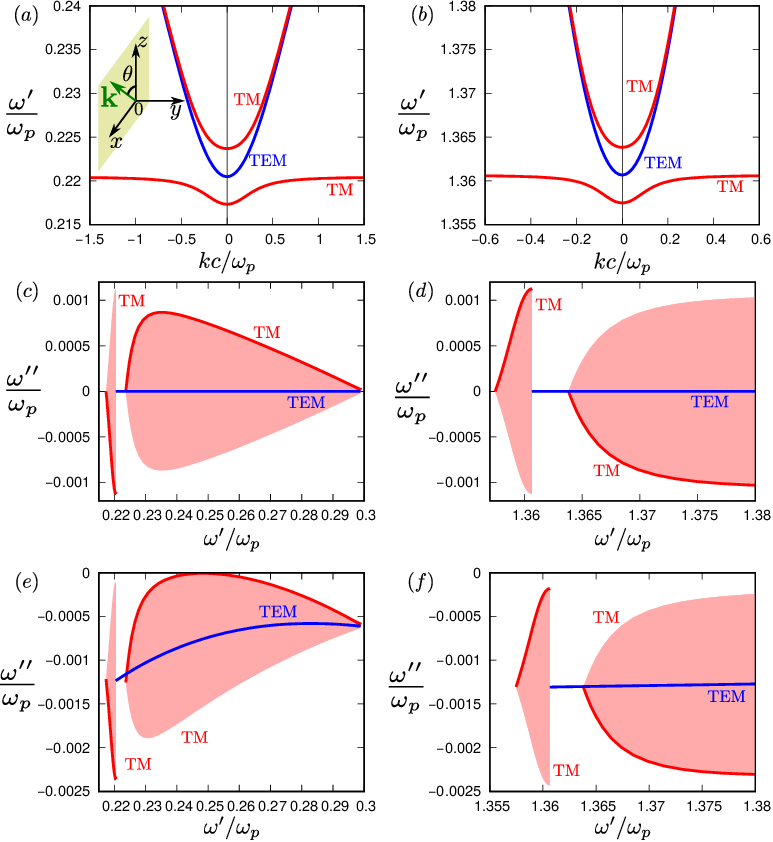}
       \caption{Dispersion of the plane wave modes $\omega=\omega'+i \omega''$ vs. $\vec{k}$ for a real-valued wave vector and propagation in the $xoz$ plane. The medium is described by the permittivity tensor \eqref{E:epsilon_example}. (a) and (b): $\omega'$ as a function of $k$ near the edge the low-frequency (high-frequency) branches, respectively, for $\theta=\pi/4$. These two plots are nearly insensitive to the values of $\theta$, $\Gamma$ and $\gamma$. The wavevector in the $xoz$ plane is represented in the inset of (a). (c) and (d): Projection of the band diagram in the complex plane for the low-frequency (high-frequency) bands, respectively, ignoring collisions ($\Gamma=\gamma=0^+$). (e) and (f): Similar to (c) and (d) but for $\Gamma=3.85\cdot  10^{-3} \omega_p$ and $\gamma=1.232\cdot  10^{-3}\omega_p$.
       For TM waves, the solid red curves in (c)--(f) represent the wave dispersion for $\theta=\pi/4$ and the red shaded regions represent the frequency locus for all angles $\theta$. 
       In all the plots, $\omega_\text{b}=0.9 \omega_p$, $\omega_0=0.3\omega_p$ and  $\eps_0 v_{0x} a_{zxx}=0.01/ \omega_p$.
       }
\label{fig:band_diagram_xoz_plane}
\end{figure*}

When the coupling is nonzero ($\eps_c \neq 0$), the material becomes anisotropic.
First, we consider wave propagation in the $xoz$ plane, with the wavevector $\vec{k}=k\left( \sin(\theta) \hat{\vec{e}}_x+ \cos(\theta) \hat{\vec{e}}_z \right)$ parameterized by an angle $\theta$ represented in the inset of Fig. \ref{fig:band_diagram_xoz_plane} (a).
In this case, the plane wave solutions of the homogeneous wave equation $\nabla \times \nabla \times  \vec{E}
=  \frac{\omega^2}{c^2} \db{\eps}   \cdot  \vec{E}$ consist of transverse magnetic (TM) and TEM modes. The TEM mode (ordinary wave) is characterized by an electric field polarized along the $y$ direction and, similar to the unbiased case, by the wavenumber $k = \frac{\omega}{c} \sqrt{\eps_d }$.
The TM mode (extraordinary wave) is characterized  by
\begin{subequations}\label{E:TM_mode}
\begin{align} 
&k  = \frac{\omega}{c} \sqrt{\eps_\text{TM} } \equiv k_\text{TM}  \\
&\eps_\text{TM}=\frac{\eps_d^2 - 2\eps_c^2 }{\eps_d^2+ \left( \cos(\theta) \sin(\theta)  \eps_c  \right)^2}\left( \eps_d- i \cos(\theta) \sin(\theta)  \eps_c  \right)  \label{E:effective_permittivity_TM} \\ 
&\vec{E}   \sim A_\text{TM} \left(\hat{\vec{e}}_x - \nu\hat{\vec{e}}_z   \right) \e{ik_\text{TM}\left( \sin(\theta) x+ \cos(\theta) z \right)}\equiv \vec{E}_\text{TM} 
\end{align} 
\end{subequations}
where $A_\text{TM}$ is a constant and $\nu=\frac{\eps_d-\eps_\text{TM}\cos(\theta)^2}{\eps_\text{TM}\cos(\theta) \sin(\theta) -i \eps_c }$ is a complex-valued function that depends on $\omega$ and $\theta$.
For propagation in the $xoz$ plane, the TEM and TM eigenmodes are orthogonal.
It is seen from equation \eqref{E:effective_permittivity_TM} that even in the absence of collisions (i.e. $\eps_d$ and $\eps_c$ real), the effective permittivity of the TM mode $\eps_\text{TM}$ is complex valued, with an imaginary part that is maximized/minimized for $\theta=\mp \pi/4$ and $\theta=\pm 3\pi/4$. Thus in the absence of collisions, depending on the value of $\theta$, $\eps_\text{TM}$ can have either a positive or negative imaginary part implying that the material may either dissipate or amplify an electromagnetic wave. An intriguing consequence is that an antenna embedded in this material would radiate a beam that amplifies within a specific angular region while being suppressed in other directions.

In the following, we consider plane wave solutions with a real-valued wave vector $\vec{k}$. This models a spatially periodic system or a system closed on itself (e.g., a closed cavity). As $\eps_\text{TM}$ has a nonzero imaginary part, the corresponding oscillation frequencies $\omega=\omega'+i \omega''$ are complex-valued.
For the time convention $\e{-i\omega t}$, an oscillation with $\omega''<0$ indicates a relaxation due to material absorption, whereas an oscillation with $\omega''>0$ represents an instability due to material gain.

The dispersion $\omega'$ vs. $k$ of the electrically-biased material is represented in Figs. \ref{fig:band_diagram_xoz_plane}(a) and (b)  for $\theta=\pi/4$.
The dispersion of the TEM modes (ordinary wave) is the same as in the unbiased case (see Fig. \ref{fig:band_diagram_no_coupling}). On the other hand, for each branch [low-frequency bands in Fig. \ref{fig:band_diagram_xoz_plane}(a) and high-frequency bands in Fig. \ref{fig:band_diagram_xoz_plane}(b)] the TM waves split into two sub-bands. The TM waves can be understood as resulting from the hybridization, due to $\eps_c \neq 0$, between  one of the TEM waves of the unbiased system with the longitudinal wave.

In Figs. \ref{fig:band_diagram_xoz_plane} (c)--(f) we represent the projected band diagram in the complex plane, corresponding to the locus of $\omega'(\vec{k})+i \omega''(\vec{k})$ across all angles $\theta$.
When collisions are ignored [Fig. \ref{fig:band_diagram_xoz_plane} (c) and (d)], the directions of propagation are evenly divided between stable excitations ($\omega''<0$) and unstable modes ($\omega''>0$). The complex spectrum exhibits mirror symmetry with respect to the real-frequency axis.
Therefore, in these conditions the material response is unstable, suggesting potential applications for lasing. 

The effect of collisions can stabilize the material response and ensure that the frequency spectrum is fully confined to the lower-half frequency plane for all $k$ real-valued and all directions of propagation. This is illustrated in  Figs. \ref{fig:band_diagram_xoz_plane} (e) and (f), where the collisions strength was adjusted to the stability threshold, resulting in  all the modes having $\omega''\leq 0$.

\subsection{Indefinite gain response}
\label{Sec:indefinite_gain}

In order to explain the directional dependence of the gain response of the material, next we characterize the tensor $\db{\eps}'' \equiv \frac{\db{\eps}   - \db{\eps}^\dag }{2i}$, which determines the power exchanged between the material and the wave \cite{lannebere_nonreciprocal_2022,lannebere_chiral_2025,lannebere_symmetry_2025}. For the model of Eq. \eqref{E:epsilon_example}, it is given by $\db{\eps}''=\mathrm{Im}\left(\eps_d\right) \vec{1}_{3\times3} + \db{\chi}_\text{NH} $ where 
\begin{align}  \label{E:chi_NH}
\db{\chi}_\text{NH} &=  \begin{pmatrix} 0 & 0 & \frac{1}{2}   \mathrm{Re}\left(\eps_c\right)- i \frac{3}{2} \mathrm{Im}\left(\eps_c\right) \\ 0&0&0\\  \frac{1}{2}   \mathrm{Re}\left(\eps_c\right)+ i \frac{3}{2} \mathrm{Im}\left(\eps_c\right) &0&0 \end{pmatrix}.
\end{align}
For a standard passive material, all the eigenvalues of $\db{\eps}''$ are required to be positive to guarantee that the energy of the wave is irreversibly dissipated in the material. Interestingly, in our problem 
$\db{\chi}_\text{NH}$ is an indefinite matrix, meaning that it has both positive and negative eigenvalues. Specifically, its nontrivial eigenvalues are $\pm \frac{1}{2} |\alpha_0|$, with $\alpha_0=\mathrm{Re}(\eps_c)-3 i \mathrm{Im}(\eps_c)$, and the corresponding eigenvectors are  $\pm \frac{\alpha_0 }{|\alpha_0|} \hat{\vec{e}}_x + \hat{\vec{e}}_z$. 
The positive eigenvalue corresponds to an electro-optic response that results in increased dissipation, whereas the negative eigenvalue  to an electro-optic response that results in reduced dissipation. 

As $\db{\chi}_\text{NH}$ trivially commutes with the response of the unbiased material (a scalar), the eigenvalues of $\db{\eps}''$ are
$\lambda_\pm\equiv\mathrm{Im}(\eps_d) \pm \frac{1}{2} |\alpha_0|$, corresponding to the eigenvectors $\pm \frac{\alpha_0 }{|\alpha_0|} \hat{\vec{e}}_x + \hat{\vec{e}}_z$, and $\lambda_0\equiv\mathrm{Im}(\eps_d)$ corresponding to the eigenvector $\hat{\vec{e}}_y$.
Thus, in the presence of an electric bias the level of dissipation depends on the wave polarization in the material. In our system, the parameter $\alpha_0$ is predominantly real-valued. Thereby, the polarizations that activate the maximum (minimum) dissipation are such that $\vec{E} \sim s \hat{\vec{e}}_x + \hat{\vec{e}}_z$ and $\vec{E} \sim -s \hat{\vec{e}}_x + \hat{\vec{e}}_z$, with $s = \text{sgn} \left( \text {Re}\left\{\eps_c \right\} \right)$. This explains why the maximum dissipation/gain occurs for angles $\theta= \mp \pi/4$: for such directions of propagation the electric field of the TM wave is roughly oriented along $\pm \pi/4$, matching closely one of the eigenvectors of $\db{\eps}''$.

Eventually, when the bias velocity is strong enough such that in some frequency range $ \frac{1}{2} |\alpha_0|>\mathrm{Im}(\eps_d)$, the smallest eigenvalue $\lambda_-$ can become negative. In such a case, the material exhibits an indefinite gain response \cite{lannebere_chiral_2025}, as it can provide both dissipative and gain interactions, depending on the wave polarization.
When $\lambda_-<0$, the electromagnetic waves polarized as $-\frac{\alpha_0 }{|\alpha_0|} \hat{\vec{e}}_x + \hat{\vec{e}}_z$ are amplified. 

\begin{figure*}[!ht]
\centering
\includegraphics[width=0.95\linewidth]{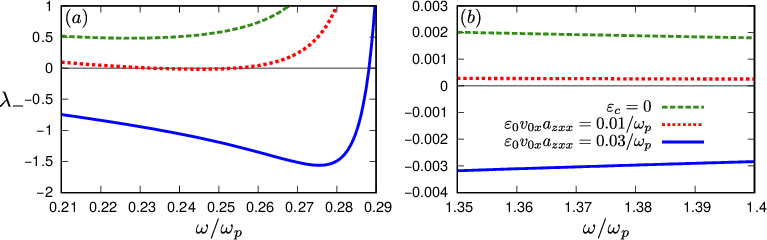}
       \caption{Plot of $\lambda_-$, the smallest eigenvalue of $\db{\eps}''$, for a frequency range corresponding to (a) the low-frequency band and (b) the high-frequency band. The green dashed curve represents the unbiased case. The red dotted curve corresponds to $\eps_0 v_{0x} a_{zxx}=0.01/ \omega_p$ and the blue solid line to $\eps_0 v_{0x} a_{zxx}=0.03/ \omega_p$. The remaining simulation parameters are the same as in Fig. \ref{fig:band_diagram_xoz_plane}, with the effect of collisions included.
       }
\label{fig:eigenvalues_eps''}
\end{figure*}
Figure \ref{fig:eigenvalues_eps''} depicts $\lambda_-$ as a function of frequency for different values of the bias velocity.
In the unbiased case (green dashed curve), $\lambda_-=\mathrm{Im}(\eps_d)$ is positive for all frequencies, consistent with the passivity of the material. When the bias velocity is increased beyond some threshold, $\lambda_-$  can reach negative values, indicating that the associated eigenpolarization can be amplified by the material. As seen, the electro-optic response is  stronger in the low-frequency band, where $\lambda_-$ is more negative than in the high-frequency band. In particular, for the case corresponding to the stability threshold of the closed system (red dotted curve), $\lambda_-$ only reaches negative values in the low-frequency band whereas it is positive everywhere in the high-frequency band. It is interesting to note that the system can exhibit gain ($\lambda_-<0$) even in the regime where the closed system is stable ($\omega''\le 0$ for all real-valued wave vectors). This occurs because the eigenpolarization of the TM-polarized plane wave, $\vec{E}_\text{TM}$, is not close enough to $-\frac{\alpha_0 }{|\alpha_0|} \hat{\vec{e}}_x + \hat{\vec{e}}_z$, the eigenvector associated with $\lambda_-$ that triggers the material gain. The transverse optical phonon frequency $\omega_0$ of natural materials is typically in the range of the tens of THz \cite{kittel_introduction_2004}. Thus, the NHEO effect should be the strongest for frequencies in the THz range.

In the low-frequency band, the frequency that gives the most negative $\lambda_-$, i.e. the strongest gain, depends on the bias strength and is close to $\omega=0.244\omega_p$ ($\omega=0.275\omega_p$) for the red dotted curve (blue solid line). 

\subsection{Traveling wave amplifier}

Next, we demonstrate that the electrically biased material can be used as a traveling wave amplifier. To this end, we examine the Poynting vector of a TM wave. Different from subsection \ref{Sec:stability}, we consider plane wave solutions with a real-valued frequency $\omega$ and a complex-valued wave vector $k_\text{TM}=k_\text{TM}^\prime+ik_\text{TM}^{\prime\prime}$. This models for example the wave propagation in an open system of finite dimensions. 

The Poynting vector $\vec{S}_\text{TM}=\frac{1}{2} \mathrm{Re}\left\{\vec{E}_\text{TM} \times \vec{H}_\text{TM}^\ast \right\} $ with $\vec{H}_\text{TM}=\frac{1}{i \omega \mu_0} \nabla \times \vec{E}_\text{TM}$, is given explicitly by:
\begin{align}\label{E:poynting_TM}
\vec{S}_\text{TM}&= \frac{|A_\text{TM}|^2}{ 2\omega \mu_0}  \mathrm{Re}\left\{         k_\text{TM} \left[ \cos(\theta)  + \nu \sin(\theta) \right] \left( \nu^\ast \hat{\vec{e}}_x +\hat{\vec{e}}_z \right)\right\} \e{-2 k_\text{TM}^{\prime\prime} \left( \sin(\theta) x+ \cos(\theta) z \right)}.
\end{align}
This vector exhibits different behavior depending on the sign of the imaginary part of $k_\text{TM}$. When $k_\text{TM}^{\prime\prime}$ is positive, the Poynting vector experiences exponential decay, corresponding to a wave attenuated by the material, whereas when $k_\text{TM}^{\prime\prime}<0$, the Poynting vector grows exponentially, corresponding to a wave amplified by the material. The threshold electric bias required to have amplification ($k_\text{TM}^{\prime\prime}<0$) is the same as the stability threshold for the closed system discussed in subsection \ref{Sec:stability}. It is important to note that while such a system may be unstable under closed boundary conditions, a system with open boundaries—lacking the feedback loop provided by periodic boundary conditions—can, in principle, remain stable.

Note that similar to passive anisotropic media \cite{kong_electromagnetic_1986}, the Poynting vector $\vec{S}_\text{TM}$ is not parallel to the direction of propagation $\vec{k}$. 
Provided $\left| \varepsilon_c \right| \ll \left| \varepsilon_d \right|$, so that the material response is quasi-isotropic,  the angle between both vectors is quite small. For example, in the numerical example discussed below the angle is typically smaller than 1\degree.

\begin{figure*}[!ht]
\centering
\includegraphics[width=\linewidth]{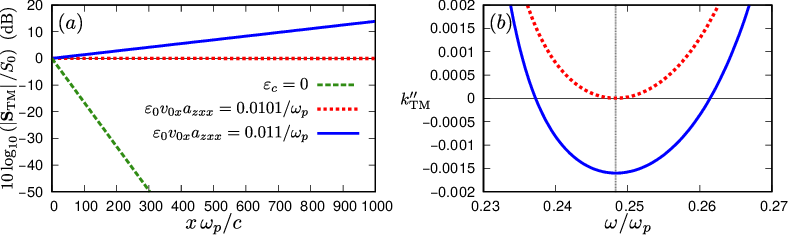}
       \caption{ (a) Amplitude of the Poynting vector $\vec{S}_\text{TM}$ expressed in dB as a function of the propagation distance for $x=z$, $\theta=\pi/4$ and $\omega= 0.2483 \omega_p$. The green dashed curve represents the unbiased case. The bias strength for the other two curves is indicated in the inset. The remaining simulation parameters are the same as in Fig. \ref{fig:band_diagram_xoz_plane}, with the effect of collisions included. $S_0$ is the Poynting vector amplitude at the input $S_0=\left|\vec{S}_\text{TM}(x=z=0)\right|$. (b) Imaginary part of $k_\text{TM}$ as a function of the frequency in the low-frequency band for the same parameters as in (a). The curve for the unbiased system is outside the  plot range. The vertical gray dotted line marks the frequency used in (a).}
\label{fig:travelling_wave_amplifier}
\end{figure*}

To illustrate the gain effect, we represent in Fig. \ref{fig:travelling_wave_amplifier} (a) the amplitude of the Poynting vector $\vec{S}_\text{TM}$ [Eq.\eqref{E:poynting_TM}] in dB as a function of the propagation distance for  different values of the electric bias. 
The frequency and the direction of propagation ($\theta=\pi/4$) are selected to ensure maximum sensitivity to the gain effect. In the absence of bias, $k_\text{TM}^{\prime\prime}$ is positive due to the loss of the unbiased system (not shown), resulting in a Poynting vector  decaying exponentially with the propagation distance (green dashed curve). As the strength of the electric bias  increases, $k_\text{TM}^{\prime\prime}$ and the decay rate progressively decrease until reaching the point where $k_\text{TM}^{\prime\prime}=0$ [minimum of the red dotted curve in Fig. \ref{fig:travelling_wave_amplifier} (b)]. At this point, there is no net energy absorption as illustrated with the red dotted curve in Fig. \ref{fig:travelling_wave_amplifier} (a). The wave dissipation due to collisions is exactly compensated by the gain provided by the electro-optic effect, allowing the wave to propagate through the material without attenuation.

For bias strengths above the stability threshold of the closed system, $k_\text{TM}^{\prime\prime}$ becomes negative in some frequency range in the low-frequency band, as shown by the solid blue line in Fig. \ref{fig:travelling_wave_amplifier} (b). The corresponding Poynting vector grows exponentially in this range as shown in Fig. \ref{fig:travelling_wave_amplifier}(a) (solid blue line). In this scenario, the system can function as a traveling wave amplifier. 
 
\subsection{Faraday-type rotation and electromagnetic isolation}
\label{Sec:Faraday}

\begin{figure*}[!ht]
\centering
\includegraphics[width=0.85\linewidth]{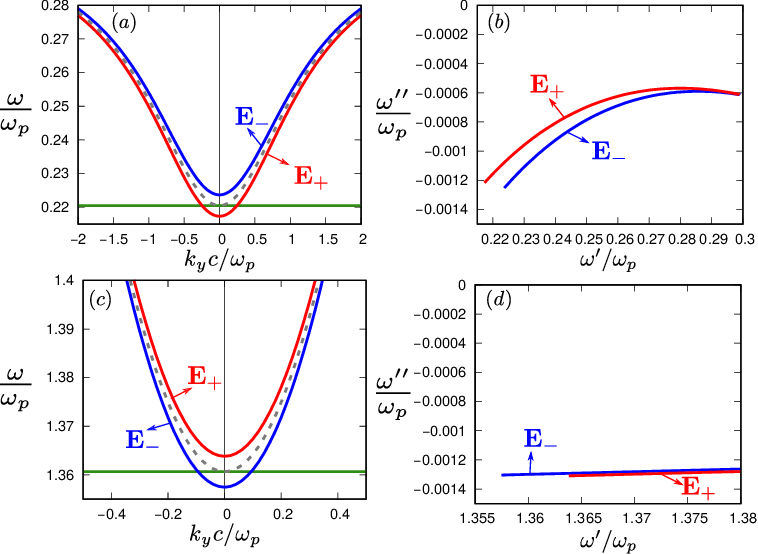}
       \caption{Dispersion of the plane wave modes $\omega=\omega'+i \omega''$ for propagation along the $y$ direction and a real-valued wave vector $k_y$. (a) and (c): $\omega'$ as a function of $k_y$ near the edge of the  low-frequency band (high-frequency band), respectively. (b) and (d): Projected band structure in the complex plane for the low-frequency band (high-frequency band), respectively, and propagation along $y$.
       The red and blue solid lines represent the dispersions of the modes associated with $\vec{E}_+$ and $\vec{E}_-$, respectively. In (a) and (c) the horizontal green lines represent the longitudinal modes, whereas the dashed grey line is the TEM mode in the unbiased case ($\eps_c=0$). The simulation parameters are the same as in Fig. \ref{fig:band_diagram_xoz_plane}, with the effect of collisions included.}
\label{fig:band_diagram_azxx}
\end{figure*}

Next, we focus on wave propagation along the $y$ direction. 
The plane wave solutions  with wave vector $\vec{k}=k_y\hat{\vec{e}}_y$
consist of a longitudinal mode and two nondegenerate TEM modes. The longitudinal mode dispersion is the same as in the unbiased system. The two TEM modes, labeled $\pm$ are characterized by
\begin{subequations}\label{E:transverse_modes}
\begin{align}
&k_y = \frac{\omega}{c} \sqrt{\eps_\pm } \label{E:k_plus_minus} \equiv k_\pm\\ 
&\eps_\pm(\omega)=\eps_d \pm \sqrt{2}\eps_{c} \label{E:epsilon_plus_minus}\\
&\vec{E} \sim \frac{1}{\sqrt{3}} \left( \hat{\vec{e}}_x \pm i \sqrt{2}     \hat{\vec{e}}_z \right) e^{i k_{\pm} y}\equiv\vec{E}_\pm
\end{align} 
\end{subequations}
The dispersion diagram for propagation along $y$ is depicted in Fig. \ref{fig:band_diagram_azxx}.
As seen, the effect of the electric bias ($\eps_c \neq 0$) is to lift the degeneracy between the two TEM modes of the unbiased material. The projected band structure in the complex plane [Fig. \ref{fig:band_diagram_azxx} (b) and (d)] shows that the spectrum is fully contained in the lower-half frequency plane, confirming that the system is stable for this bias strength. 
The eigenvectors $\vec{E}_\pm$ associated with the nondegenerate TEM modes are elliptically polarized, and have opposite handedness.  Remarkably, as soon as $\eps_c \neq 0$ the two eigenvectors cease to be orthogonal: $\vec{E}_+ \cdot \vec{E}_-^\ast \neq 0$. This feature is a consequence of the non-Hermitian response of the system \cite{lannebere_nonreciprocal_2022}. 

The structure of the eigenmodes resembles the eigenpolarizations observed in a magnetized plasma when waves propagate along the magnetic bias direction. However, in this case, the waves are elliptically polarized instead of being circularly polarized.  Next, we exploit this similarity to develop a Faraday-type rotator.

To this end, we derived the transmission matrix $\db{T}$ that relates the transverse components of the incident $\vec{E}_t^\text{inc}=\left(E_{x}^\text{inc} \; E_{z}^\text{inc} \right)^T$ and transmitted $\vec{E}_t^\text{tr}=\left(E_{x}^\text{tr} \; E_{z}^\text{tr} \right)^T$ electric fields as $
 \vec{E}_t^\text{tr} =\db{T} \cdot \vec{E}_t^\text{inc} $.
It is shown in appendix \ref{sec:relection_transmission_slab} that for normal incidence (incident wave propagates along $+y$) the transmission matrix for a material slab (infinitely extended along $x$ and $z$) of thickness $d$ is 
\begin{align}  \label{E:T_matrix}
\db{T} = \frac{1}{2}\begin{pmatrix}
     \beta_+ +\beta_-   & \frac{-i}{ \sqrt{2}} \left[ \beta_+ -\beta_- \right]  \\
 i \sqrt{2} \left[ \beta_+ -\beta_- \right] &   \beta_+ +\beta_- 
\end{pmatrix},
\end{align}
where $\beta_j=\left[ \cos (k_j d)-i\frac{ (\eps_j +\eps_\text{ext}) }{2 \sqrt{\eps_j} \sqrt{\eps_\text{ext}}}\sin (k_j d)\right]^{-1}$ with $j=+$ or $-$, $\eps_\text{ext}$ the permittivity of the surrounding dielectric (e.g., air) and the remaining  parameters defined as in Eq. \eqref{E:transverse_modes}. 

\begin{figure*}[!ht]
\centering
\includegraphics[width=0.4\linewidth]{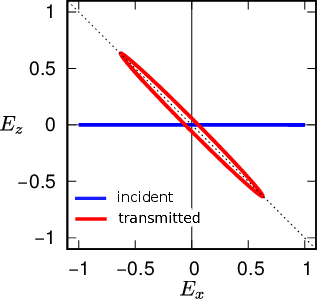}
       \caption{Trajectory of the electric field over an optical cycle (polarization curve) for the incident ($\vec{E}_t^\text{inc}$) and the transmitted ($\vec{E}_t^\text{tr}$) waves. The incident wave illuminates a slab of thickness $d\approx 0.246 \lambda_0$ with $\lambda_0=2\pi c/\omega$ the vacuum wavelength. The slab stands in air ($\eps_\text{ext}=1$) and the operating frequency is $\omega=0.2277 \omega_p$. The collision frequencies are $\Gamma=3.85\cdot  10^{-3} \omega_p$, $\gamma=1.232\cdot  10^{-3}\omega_p$ and the remaining structural parameters are the same as in Fig. \ref{fig:band_diagram_xoz_plane}.}
\label{fig:transmission_slab}
\end{figure*}

Figure \ref{fig:transmission_slab} depicts the polarization curve for the wave transmitted through a material slab with thickness $d\approx 0.246\lambda_0$ (red curve), with the incident wave linearly polarized along the $x$ direction (polarization curve in blue).
To replicate the functionality of standard Faraday isolators, the set of parameters was chosen so that the transmitted wave undergoes a rotation of approximately $45\degree$. As seen, the transmitted wave is slightly elliptically polarized.  Remarkably, owing to the non-Hermitian response and the subwavelength thickness of the slab, the transmission level remains high, even though a reasonable level of loss (due to collisions) is taken into account. The transmission matrix \eqref{E:T_matrix} is not invariant under rotations about the $y$-axis. Thus, the level of rotation may depend somewhat on the orientation of the incident electric field. Moreover $\db{T}$ is independent of the direction of propagation of the incoming wave.  Consequently, it is feasible to design a compact electromagnetic isolator with low insertion loss, simply by inserting the subwavelength electrically-biased material slab in between two linear polarizers rotated by an angle close to $45\degree$, analogous to the configuration of a standard Faraday isolator. 

\subsection{Power beating}

Next, we examine the Poynting vector for a superposition of plane waves. The electric field for a superposition of $+$ and $-$ modes propagating along the $+y$ direction is given by 
\begin{align}  \label{E:E_field_superposition}
\vec{E}= \frac{A_+}{\sqrt{3}} \left( \hat{\vec{e}}_x + i \sqrt{2}     \hat{\vec{e}}_z \right) \e{i k_+ y} 
+  \frac{A_-}{\sqrt{3}} \left( \hat{\vec{e}}_x - i \sqrt{2}     \hat{\vec{e}}_z \right) \e{i k_- y} 
\end{align}
where $A_+$ and $A_-$ are complex coefficients. The corresponding magnetic field is $\vec{H}=\frac{1}{i \omega \mu_0} \hat{\vec{e}}_y \times \partial_y \vec{E}$. 
After some algebra, one can show that the time-averaged Poynting vector $\vec{S}=\frac{1}{2} \mathrm{Re}\left(\vec{E} \times \vec{H}^\ast \right) $ is given by:
\begin{align} \label{E:poynting}
\vec{S} = \frac{1}{2\omega \mu _0} \text{ Re} \left\{ {k_ + ^*{{\left| {{A_ + }} \right|}^2}{e^{ - 2{k^{\prime\prime}_{+} }y}} + k_ - ^*{{\left| {{A_ - }} \right|}^2}{e^{ - 2{k^{\prime\prime}_{-} }y}} - \frac{1}{3}\left( {k_ + ^{} + k_ - ^*} \right){A_ + }A_ - ^*{e^{i\left( {{k_ + } - k_ - ^*} \right)y}}} \right\}\hat{\vec{e}}_y.
\end{align}
 %
%
We write the wave numbers in terms of real and imaginary parts: $k_\pm = k^\prime_\pm + i k^{\prime\prime}_\pm$. As seen, similar to what happens in other non-Hermitian systems \cite{lannebere_nonreciprocal_2022,buddhiraju_nonreciprocal_2020,hadad_possibility_2020,fernandes_exceptional_2024}, the Poynting vector is the sum of the power transported by the $+$ and $-$ waves, plus an interference term that oscillates with spatial frequency  $k'_+ - k'_-$. 

\begin{figure*}[!ht]
\centering
\includegraphics[width=0.55\linewidth]{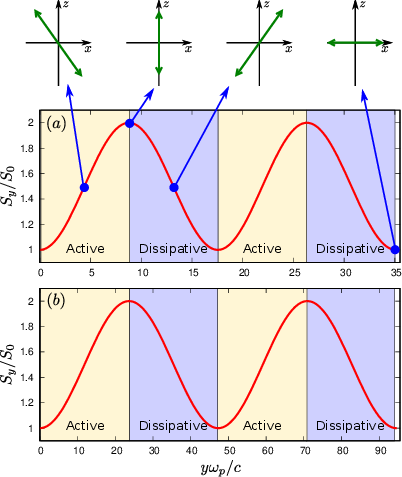}
       \caption{ Normalized Poynting vector as a function of the propagation distance $y$ for  (a) $\omega\approx 0.224 \omega_p$ at the edge of the low-frequency band associated with the $-$ mode. (b) $\omega \approx 1.364 \omega_p$ at the edge of the high-frequency band associated with the $+$ mode. The parameters of the simulation are $A_+=A_-=1$, $\Gamma=\gamma \to 0^+$ and the remaining parameters are as in Fig. \ref{fig:band_diagram_xoz_plane}. The electric field polarizations associated with the points where the curve slope  either vanishes or is a maximum/minimum are represented in (a). The field polarizations in case (b) are nearly identical.}
\label{fig:poynting}
\end{figure*}

To illustrate this behavior, we represent in Fig. \ref{fig:poynting} the $y$ component of the Poynting vector [Eq.\eqref{E:poynting}] as a function of the propagation distance. For clarity, in this plot we use $\gamma = \Gamma = 0^+$, i.e., we assume weak dissipation. Thus, the two propagation constants $k_{\pm}$ become real-valued.  

The two operating frequencies of Fig. \ref{fig:poynting} are coincident with the cutoff frequencies of the $-$ ($+$) mode in the low (high)-frequency band, respectively (see Fig. \ref{fig:band_diagram_azxx}). Such a choice ensures that $|k_+ - k_-|$ is maximized in each band and provides the highest rate of gain/loss per unit of length. As seen, similar to an ideal MOSFET-metamaterial \cite{lannebere_nonreciprocal_2022}, 
our system alternates between active and dissipative states depending on the wave polarization. The wave amplification and attenuation are a result of the energy exchanged between the dynamic fields and the static current. This exchange is rooted in nonlinear mechanisms, akin to the operation of a field-effect transistor \cite{fernandes_exceptional_2024}.

As shown in Sec. \ref{Sec:indefinite_gain}, in the limit $\gamma = \Gamma = 0^+$ the wave amplification/absorption is maximized for two orthogonal linear polarizations ($\theta = \pm 45 \degree$). In the example of Fig. \ref{fig:poynting}, we consider a wave with $A_+=A_-$. In this scenario, the polarization remains linear during propagation. The polarization state is continuously rotated as the wave propagates in the material, in agreement with the discussion in subsection \ref{Sec:Faraday}.
In general, as soon as $|A_+| \neq |A_-|$, the polarization in the material becomes elliptical.  As $|k_+ - k_-|$ is larger for the lower-band, the amplification/absorption is stronger in this band. Moreover, the spatial oscillation period is shorter for this band. 
\begin{figure*}[!ht]
\centering
\includegraphics[width=\linewidth]{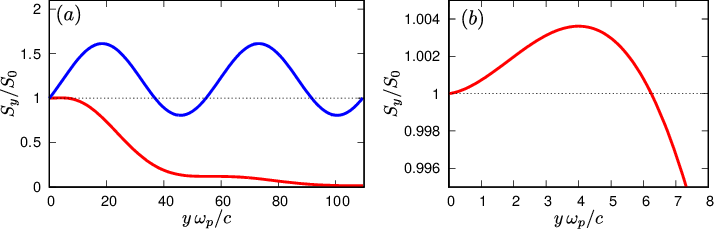}
       \caption{(a) Normalized Poynting vector as a function of the propagation distance $y$ for $\omega= 0.244 \omega_p$. The blue curve corresponds to the lossless case $\Gamma=\gamma=0$ and the red curve corresponds to $\Gamma=3.85\cdot  10^{-3} \omega_p$ and $\gamma=1.232\cdot  10^{-3}\omega_p$. (b) Zoom in of the red curve in the region close to $y=0$. In all the plots $A_+=1$, $A_-\approx \e{-i 1.023}$ and the rest of the parameters are as in Fig. \ref{fig:band_diagram_xoz_plane}. The horizontal dotted line marks the Poynting vector amplitude at the input $S_y=S_0$.}
\label{fig:poynting_vs_loss_2}
\end{figure*}
In the presence of collisions, the situation is qualitatively different because $k_\pm$ are complex-valued for a real-valued frequency $\omega$. In this case, the Poynting vector exhibits exponential decay as represented  in Fig. \ref{fig:poynting_vs_loss_2}  for a value of the collisions that ensure a stable behavior of the closed system.
As seen, despite the exponential decay, the system can still provide a small amplification, in agreement with the results of Sec. \ref{Sec:indefinite_gain}. Here, the parameters were chosen to maximize the gain so that the system is operated at the minimum of the red dotted curve in Fig.\ref{fig:band_diagram_azxx} (a). For the sake of comparison, the Poynting vector in the absence of collisions was also represented in the figure. 

To conclude, it is important to mention that the effects studied in this section are not only achievable with a coupling proportional to $a_{zxx}$, but also with the other couplings described in Eq. \eqref{E:linearized_permittivity}.

\section{Conclusion}
\label{sec:conclusion}

In conclusion, this paper introduces a novel mechanism to implement the NHEO effect, achieving optical gain through nonlinear interactions between free and bound electrons. This mechanism operates independently of the anomalous velocity of Bloch electrons and the Berry curvature dipole. Our phenomenological model describes both the nonlinearities arising from the anomalous velocity, as well as nonlinearities arising from the interband effects, which are not captured by the usual Boltzmann approximation. The model shows that in low-symmetry conductors under non-equilibrium conditions, the linearized electromagnetic response becomes not only nonreciprocal but also capable of exhibiting polarization-dependent gain, even in the absence of the anomalous velocity contribution.

Our model ensures global passivity and adheres to the principles of microscopic reversibility. By considering the specific spatial symmetries of the model, we can significantly narrow down the permissible forms of nonlinear coupling. A notable aspect of the NHEO response is that it necessitates broken inversion symmetry. We demonstrate that an electrically biased material with the symmetries of the \textit{2mm} point group  can exhibit a robust nonreciprocal and non-Hermitian response.
We identify (natural) materials belonging to the class of polar/ferroelectric metals with \textit{2mm} point group symmetry, such as for example WTe$_2$, as potential  candidates to implement this effect experimentally. 

Furthermore, we analyzed the wave propagation  within these systems and evaluated the application of our electrically biased material in devices such as kinetic Faraday rotators, Faraday isolators, and traveling wave amplifiers. In addition, the strength of the electric bias can be precisely adjusted to compensate for material dissipation, facilitating the realization of robust waveguides for nanophotonic applications. Our findings open new pathways for the use of electrically biased systems in photonic circuits, presenting a potentially interesting alternative to traditional methods reliant on magnetic effects or optical pumping. 

\section*{Acknowledgments}
This work was partially funded by the
Institution of Engineering and Technology (IET) under the A F Harvey
Research Prize 2018, by the Simons Foundation under the award SFI-MPS-EWP-00008530-04 (N.E.) and award SFI-MPS-EWP-00008530-10 (M.S.) (Simons Collaboration in Mathematics and Physics, "Harnessing
Universal Symmetry Concepts for Extreme Wave Phenomena") and by FCT/MECI through national funds and when applicable co-funded EU funds under UID/50008: Instituto de Telecomunica\c{c}\~{o}es. S.L. acknowledges FCT and IT-Coimbra for the research financial support with reference DL
57/2016/CP1353/CT000.  

\appendix

\section{Linearization of nonlinear conservative systems described by a Lagrangian} \label{sec:Lagrangian}
In this Appendix, we demonstrate that the linearization of nonlinear conservative systems, described by a Lagrangian and subjected to a time independent bias (equilibrium situation), results invariably in a Hermitian response.

We consider a general Lagrangian system in which the electromagnetic fields interact with matter. According to the classical theory of fields \cite{landau_classical_1980}, the action $S$ in this system is expressed as follows:
\begin{align}\label{E:generalized_action}
S= \int_{t_1}^{t_2}  \int \mathcal{L}_\text{NL}\left(\vec{q},\dot{\vec{q}},\partial_x {\vec{q}} \right)  dx  dt
\end{align}
where $\partial_x=\partial/\partial x$, $\dot{\vec{q}}=\partial\vec{q}/\partial t$, $\mathcal{L}_\text{NL}$ is the Lagrangian density and $\vec{q}$ represents either the matter fields (such as the displacement of the negative charge ``cloud'' with respect to the ion background, which is proportional to the polarization vector) or the components of the electromagnetic four-potential. For simplicity of notation, we consider only one spatial dimension ($x$) in Eq. \eqref{E:generalized_action}, but the generalization to three spatial dimensions is straightforward. 

The $N$ equations of motion [$N \equiv \text{dim}(\vec{q})$] of the system  are obtained from the Euler-Lagrange equations \cite{landau_classical_1980}
\begin{align}
\frac{\partial \mathcal{L}_\text{NL}}{\partial q_i}=  \frac{\partial  }{\partial t}\left(\frac{\partial \mathcal{L}_\text{NL}}{\partial \dot{q}_i}\right)+   \frac{\partial  }{\partial x}\left(\frac{\partial \mathcal{L}_\text{NL}}{\partial (\partial_x q_i)}\right), \quad 1\leq i \leq N \label{E:gen_Euler_Lagrande_eq_NL} 
\end{align}
Except when the Lagrangian density $\mathcal{L}_\text{NL}$ is a quadratic function of $\bf{q}$, the system \eqref{E:gen_Euler_Lagrande_eq_NL} is inherently nonlinear. 

Let us suppose that the system is subject to an external, time-independent bias. The equilibrium bias point is defined by $\vec{q}=\vec{q}_0$, $\dot{\vec{q}}=\dot{\vec{q}}_0$ and $\partial_x\vec{q}=\partial_x\vec{q}_0$. 
We aim to linearize the equations of motion around this bias point. To this end, the terms $ \frac{\partial \mathcal{L}_\text{NL}}{\partial q_i}$, $ \frac{\partial \mathcal{L}_\text{NL}}{\partial \dot{q}_i}$ and $ \frac{\partial \mathcal{L}_\text{NL}}{\partial (\partial_x q_i)}$ in \eqref{E:gen_Euler_Lagrande_eq_NL} are expanded into a Taylor series. To characterize the linear dynamics it is enough to limit the expansion to the first order. For a generic function $u\left(\vec{q},\dot{\vec{q}},\partial_x {\vec{q}} \right)$  we have that:
\begin{align}
u\left(\vec{q},\dot{\vec{q}},\partial_x {\vec{q}} \right)\approx u\left(\vec{q}_0,\dot{\vec{q}}_0,\partial_x {\vec{q}_0} \right) + \sum_j \left. \frac{\partial u }{\partial q_j}  \right|_0 \delta q_j + \left. \frac{\partial u }{\partial \dot{q}_j}  \right|_0 \delta \dot{q}_j+ \left. \frac{\partial u }{\partial (\partial_x q_j)}  \right|_0 \delta (\partial_x q_j)
\end{align}
where  $\delta q_j$, $\delta \dot{q}_j$ and $\delta (\partial_x q_j)$ are respectively the $j$th component of the vectors $\vec{q}-\vec{q}_0$, $\dot{\vec{q}}-\dot{\vec{q}}_0$ and $\partial_x {\vec{q}}-\partial_x {\vec{q}}_0$. The subscript $0$ indicates that the functions are evaluated at the bias point. 
Substituting the Taylor expansions of $ \frac{\partial \mathcal{L}_\text{NL}}{\partial q_i}$, $ \frac{\partial \mathcal{L}_\text{NL}}{\partial \dot{q}_i}$ and $ \frac{\partial \mathcal{L}_\text{NL}}{\partial (\partial_x q_i)}$ in Eq. \eqref{E:gen_Euler_Lagrande_eq_NL} and taking into account that the equilibrium state satisfies the Euler-Lagrange equations, $\left. \frac{\partial \mathcal{L}_\text{NL}}{\partial q_i} \right|_{0}=\frac{\partial  }{\partial t}\left(\left. \frac{\partial \mathcal{L}_\text{NL}}{\partial \dot{q}_i} \right|_{0} \right)+ \frac{\partial  }{\partial x}\left(\left. \frac{\partial \mathcal{L}_\text{NL}}{\partial (\partial_x q_i)} \right|_{0}  \right)$, we find that the linearized equations of motion are 
\begin{align}\label{E:gen_linearized_NL_Euler_Lagrande_simple}
&  \left. \frac{\partial^2 \mathcal{L}_\text{NL}}{\partial q_j \partial q_i} \right|_0 \delta q_j + \left.\frac{\partial^2 \mathcal{L}_\text{NL}}{\partial \dot{q}_j  \partial q_i} \right|_0 \delta \dot{q}_j+ \left. \frac{\partial^2 \mathcal{L}_\text{NL}}{\partial (\partial_x q_j)\partial q_i}  \right|_0 \delta (\partial_x q_j) \nonumber \\  &=  \frac{\partial  }{\partial t}\left(   \left.  \frac{\partial^2 \mathcal{L}_\text{NL}}{\partial q_j \partial \dot{q}_i} \right|_0 \delta q_j + \left. \frac{\partial^2 \mathcal{L}_\text{NL}}{\partial \dot{q}_j \partial \dot{q}_i} \right|_0\delta \dot{q}_j+ \left. \frac{\partial^2 \mathcal{L}_\text{NL}}{\partial (\partial_x q_j) \partial \dot{q}_i} \right|_0 \delta (\partial_x q_j)\right)\nonumber \\ & \quad + \frac{\partial  }{\partial x}\left(  \left.  \frac{\partial^2 \mathcal{L}_\text{NL}}{\partial q_j \partial (\partial_x q_i)} \right|_0 \delta q_j + \left. \frac{\partial^2 \mathcal{L}_\text{NL}}{\partial \dot{q}_j \partial (\partial_x q_i)} \right|_0 \delta \dot{q}_j+ \left. \frac{\partial^2 \mathcal{L}_\text{NL}}{\partial (\partial_x q_j) \partial (\partial_x q_i)} \right|_0 \delta (\partial_x q_j)\right).  
\end{align}
The sum over repeated indices is implicit.
This set of equations describes the dynamics of the linearized system. 
Importantly, it is possible to find a quadratic Lagrangian $\mathcal{L}_\text{lin}(\vec{q},\dot{\vec{q}},\partial_x {\vec{q}} )$, 
that leads to the same set of linearized equations of motion. To do so, we simply impose that the Euler-Lagrange equations for $\mathcal{L}_\text{lin}$ (i.e. Eq. \eqref{E:gen_Euler_Lagrande_eq_NL} with $\mathcal{L}_\text{NL}$ substituted by $\mathcal{L}_\text{lin}$) yields\eqref{E:gen_linearized_NL_Euler_Lagrande_simple}. Thus, $\mathcal{L}_\text{lin}$ must satisfy the following conditions
\begin{align}
\frac{\partial \mathcal{L}_\text{lin}}{\partial q_i}&= \sum_j \left. \frac{\partial^2 \mathcal{L}_\text{NL}}{\partial q_j \partial q_i} \right|_0 \delta q_j + \left.\frac{\partial^2 \mathcal{L}_\text{NL}}{\partial \dot{q}_j  \partial q_i} \right|_0 \delta \dot{q}_j+ \left. \frac{\partial^2 \mathcal{L}_\text{NL}}{\partial (\partial_x q_j)\partial q_i}  \right|_0 \delta (\partial_x q_j) \\
\frac{\partial \mathcal{L}_\text{lin}}{\partial \dot{q}_i}&=  \sum_j \left.  \frac{\partial^2 \mathcal{L}_\text{NL}}{\partial q_j \partial \dot{q}_i} \right|_0 \delta q_j + \left. \frac{\partial^2 \mathcal{L}_\text{NL}}{\partial \dot{q}_j \partial \dot{q}_i} \right|_0\delta \dot{q}_j+ \left. \frac{\partial^2 \mathcal{L}_\text{NL}}{\partial (\partial_x q_j) \partial \dot{q}_i} \right|_0 \delta (\partial_x q_j)\\
\frac{\partial \mathcal{L}_\text{lin}}{\partial (\partial_xq_i)}&=  \sum_j \left.  \frac{\partial^2 \mathcal{L}_\text{NL}}{\partial q_j \partial (\partial_x q_i)} \right|_0 \delta q_j + \left. \frac{\partial^2 \mathcal{L}_\text{NL}}{\partial \dot{q}_j \partial (\partial_x q_i)} \right|_0 \delta \dot{q}_j+ \left. \frac{\partial^2 \mathcal{L}_\text{NL}}{\partial (\partial_x q_j) \partial (\partial_x q_i)} \right|_0 \delta (\partial_x q_j)
\end{align}
Evidently, the Lagrangian density $\mathcal{L}_\text{lin}$ is given by
\begin{align}
\mathcal{L}_\text{lin} & =    \sum_{ij} \frac{1}{2} \left( \left.  \frac{\partial^2 \mathcal{L}_\text{NL}}{\partial q_j \partial q_i} \right|_0 \delta q_j \delta q_i + \left.  \frac{\partial^2 \mathcal{L}_\text{NL}}{\partial \dot{q}_j \partial \dot{q}_i}  \right|_0 \delta \dot{q}_j \delta \dot{q}_i +   \left. \frac{\partial^2 \mathcal{L}_\text{NL}}{\partial (\partial_x q_j) \partial (\partial_x q_i)} \right|_0 \delta (\partial_x q_j)\delta (\partial_x q_i) \right) \nonumber \\ & \quad  + \sum_{ij} \left. \frac{\partial^2 \mathcal{L}_\text{NL}}{\partial \dot{q}_j \partial q_i} \right|_0 \delta \dot{q}_j\delta q_i   +    \left.  \frac{\partial^2 \mathcal{L}_\text{NL}}{\partial q_j \partial (\partial_x q_i)} \right|_0 \delta q_j \delta (\partial_x q_i) + \left. \frac{\partial^2 \mathcal{L}_\text{NL}}{\partial \dot{q}_j \partial (\partial_x q_i)} \right|_0 \delta \dot{q}_j\delta (\partial_x q_i)
\end{align}
As seen, $\mathcal{L}_\text{lin}$ is a quadratic polynomial of the three variables $\vec{q}$, $\dot{\vec{q}}$ and $\partial_x {\vec{q}}$. For a bias point that is time independent, i.e. in an equilibrium situation, it follows that $\mathcal{L}_\text{lin} $ does not depend explicitly on time. Then, it can be inferred from fundamental Lagrangian mechanics that the system described by $\mathcal{L}_\text{lin} $ is a conservative system \cite{landau_mechanics_1976}. From this, we deduce the important result that the linearized system is conservative and as a consequence it can only describe an Hermitian system with an Hermitian linearized permittivity.  
It is interesting to note that $\mathcal{L}_\text{lin}$ coincides with the second order terms of a Taylor expansion of the original Langrangian density $\mathcal{L}_\text{NL}$. 

\section{Generalized microscopic model} 
\label{Sec:ApGeneralized}
In this Appendix, we extend the model of Sec. \ref{sec:microscopic_model} to ensure that the conduction current satisfies the continuity equation. For simplicity, we ignore the effect of the ``anomalous velocity" $\left(\db{\zeta}=0\right)$, so that $\vec{p} = m \vec{v}$.

Charge conservation can be included into the model, by replacing Eqs. \eqref{E:transport_eq_simple} and \eqref{E:Lorentz} by \cite{pitaevskii_physical_2012}:
\begin{subequations}
\begin{align}  
\label{E:Aptransport_eq_simple}
    m  \left( \partial_t + \Gamma + \vec{v} \cdot \nabla  \right) \cdot \vec{v} & = q \left( \vec{E} + \db{C}_{12} \cdot \partial_t \vec{P}\right)  
  \\
  \partial_t n    + \nabla \cdot (n \vec{v} ) &=0.    \label{E:continuity_eq} \\
\label{E:Lorentz1}
m_\text{b} \left( \partial_{tt} + \gamma \partial_{t} + \sum_{j=x,y,z} \omega_{0j}^2 \hat{\vec{e}}_j\otimes \hat{\vec{e}}_j\right)  \vec{P} &= q^2 n_\text{b} \left( \vec{E} + \db{C}_{21} \cdot q n \vec{v} \right)  
\end{align}
\end{subequations}
Here, $n$ is the free electron density, which is related to the velocity through the continuity equation [Eq. \eqref{E:continuity_eq}]. It can be easily checked that provided $\db{C}_{12}^T=-\db{C}_{21}$ [Eq. \eqref{E:condition_conservative_coupling}], the fields are still constrained by a conservation law of the type $\nabla  \cdot {{\bf{S}}_{{\text{tot}}}} + {\partial _t}{W_{{\text{tot}}}} =  - Q$, with
\begin{subequations}
\begin{align}
{{\bf{S}}_{{\text{tot}}}} &= {\bf{E}} \times {\bf{H}} + \frac{1}{2}nm\left( {{\bf{v}} \cdot {\bf{v}}} \right){\bf{v}},
\\ 
\label{E:WtotAp}
{W_{{\text{tot}}}} &= {W_{{\text{EM}}}} + \frac{{{1}}}{{2}}n m{\bf{v}} \cdot {\bf{v}} + \frac{{{m_\text{b}}}}{{2{n_\text{b}}{q^2}}}
\left(
{\partial _t}{\bf{P}} \cdot {\partial _t}{\bf{P}} + \sum\limits_j {\omega _{0j}^2{\bf{P}} \cdot {{{\bf{\hat e}}}_j} \otimes {{{\bf{\hat e}}}_j}}  \cdot {\bf{P}}
\right)
,
\\ 
\label{E:QtotAp}
Q &= {{m n}\Gamma }{\bf{v}} \cdot {\bf{v}} + \gamma \frac{{{m_\text{b}}}}{{{n_\text{b}}{q^2}}}{\partial _t}{\bf{P}} \cdot {\partial _t}{\bf{P}}.
\end{align}
\end{subequations}
 
 Hence, this model also ensures that the material response remains passive ($Q>0$). Note that now both the Poynting vector ($\vec{S}_{\text{tot}}$) and the energy density ($W_{\text{tot}}$) have field and matter components.

 This refined model may be relevant for field distributions that vary fast in space, specifically when the gradient of the velocity is significant.
The velocity gradients create $\vec{k}$-dependent terms in the linearized permittivity (spatial dispersion), which can be attributed to a Doppler shift due to the electron motion. Such terms can also originate a non-Hermitian and nonreciprocal response, but are only relevant for large drift velocities and for waves that vary fast in space (e.g., surface plasmon type excitations) \cite{morgado_drift_2018, morgado_active_2021, morgado_negative_2017}. Otherwise, one can safely assume that $n(\vec{r},t) \approx  n_0 = const.$, as assumed in the main text.
 

\section{Transmission and reflection matrices for a nonmagnetic material slab under plane wave illumination} \label{sec:relection_transmission_slab}
\begin{figure*}[!ht]
\centering
\includegraphics[width=0.45\linewidth]{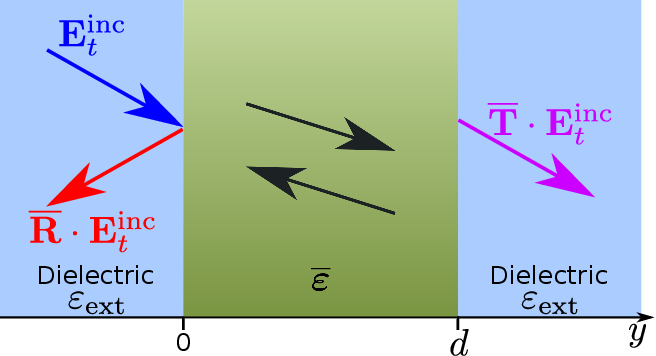}
       \caption{Geometry of a material slab with permittivity $\db{\eps}$ embedded in a dielectric of permittivity $\eps_\text{ext}$ under plane wave illumination. The slab is infinitely extended along the $x$ and $z$ directions.}
\label{fig:slab}
\end{figure*}

In this appendix, we derive the reflection and transmission matrices for a homogeneous nonmagnetic material slab subjected to plane wave incidence with an arbitrary incidence direction. The slab thickness along the $y$ direction is $d$ (see Fig. \ref{fig:slab}).
The material is described by the permittivity tensor 
\begin{align} \label{E:epsilon_MOSFET}
 \db{\eps} &= \begin{pmatrix} \eps_{xx}  & 0 &  \eps_{xz}\\ 0 & \eps_{yy} & 0 \\ \eps_{zx} & 0 & \eps_{zz}\end{pmatrix} .
 \end{align}
%

\subsection{Solutions of Maxwell equations}

In the scattering problem, the electromagnetic fields in the $x$ and $z$ directions have a spatial variation of the form $\e{i \vec{k}_t \cdot \vec{r}}$ with $\vec{k}_t=k_x \hat{\vec{e}}_x+ k_z \hat{\vec{e}}_z$. The vector $\vec{k}_t$ is determined by the spatial variation of the incident plane wave.

Then, the Maxwell equations 
\begin{align}
\nabla \times \vec{E}&= i\omega \mu_0 \vec{H} \\
\nabla \times \vec{H}&= -i\omega  \eps_0 \db{\eps} \cdot \vec{E}
\end{align}
for a material described by the permittivity tensor \eqref{E:epsilon_MOSFET} can be cast into a form similar to a Schr\"odinger's equation,
\begin{align}\label{E:propagation_mosfet}
i\frac{\partial   \vec{f}}{\partial y} 
&= \db{M} \cdot\vec{f},
\end{align}
where $\vec{f}\equiv\left( E_x \; E_z \; H_x \;  H_z \right)^T$ is the state vector and 
\begin{align} 
    \db{M}&=       \begin{pmatrix}
0 & 0 &   \frac{k_xk_z}{\omega \eps_0 \eps_{yy}} & \omega \mu_0 - \frac{k_x^2}{\omega \eps_0 \eps_{yy}}\\
0 & 0 & -\omega \mu_0 + \frac{k_z^2}{\omega \eps_0 \eps_{yy}}   &  -\frac{k_xk_z}{\omega \eps_0 \eps_{yy}}\\
 -\omega  \eps_0 \eps_{zx} - \frac{k_xk_z}{\omega \mu_0 } & -\omega  \eps_0 \eps_{zz} +  \frac{k_x^2}{\omega \mu_0 }& 0 & 0\\
\omega  \eps_0  \eps_{xx} -  \frac{k_z^2}{\omega \mu_0 }   & \omega  \eps_0  \eps_{xz} +  \frac{k_xk_z}{\omega \mu_0 }  & 0  & 0
           \end{pmatrix}.
\end{align}
The solution of Eq.\eqref{E:propagation_mosfet} is 
\begin{align} \label{E:f_field_inside}
\vec{f}(y)= \e{-iy\db{M}}\cdot \vec{f}(0)
\end{align}
where  the $4\times4$ matrix $\e{-iy\db{M}}$ is obtained by taking the exponential of the matrix of $-i y\db{M}$. Equation \eqref{E:f_field_inside} relates the state vectors $\vec{f}(y)$ and $\vec{f}(0)$ at two distinct points of the material in terms of the transfer matrix $\e{-iy\db{M}}$.

\subsection{Reflection and transmission matrices}

Next, we characterize the reflection and transmission matrices when the material slab is subjected to plane wave incidence. 
As represented in Fig. \ref{fig:slab}, the material slab interfaces are $y=0$ and $y=d$. 

Using the Maxwell equations in the dielectric region, it is feasible to relate the transverse components of the electric $\vec{E}_t=\left( E_x  \; E_z \right)^T$ and magnetic $\vec{H}_t=\left( H_x  \; H_z \right)^T$ fields associated with a plane wave through an admittance matrix \cite{morgado_single_2016,silveirinha_fluctuation_2018,latioui_lateral_2019,rappoport_engineering_2023}. The result is:
\begin{align}\label{E:relation_transverse_fields}
     \vec{H}_t^\pm   &= \pm \frac{1}{\eta_0} \vec{J} \cdot \vec{Y}_\text{ext}  \cdot \vec{E}_t^\pm,
\end{align}
where $\vec{J}=\begin{pmatrix} 0 & 1 \\-1 & 0 \end{pmatrix}$ and
\begin{align}\label{E:admittance_matrix}
 \vec{Y}_\text{ext}  =\frac{c}{ k_y  \omega  }\begin{pmatrix}     \dfrac{\omega^2\eps_\text{ext}}{c^2}- k_z^2     &    k_x  k_z  \\ k_z  k_x   & \dfrac{\omega^2\eps_\text{ext}}{c^2}-k_x^2  \end{pmatrix}
\end{align}
with $\eta_0 $ the vacuum impedance and $k_y = \sqrt{\dfrac{\omega^2\eps_\text{ext}}{c^2}-k_x^2-k_z^2}$. In Eq. \eqref{E:relation_transverse_fields} the superscript $\pm$ discriminates between plane waves propagating along the $+$ or $-y$ direction, respectively.

In the region $y<0$, the electromagnetic fields are a superposition of an incident wave and a reflected wave. We relate the reflected and incident fields at the $y=0$ interface through a reflection matrix: $ \vec{E}_t^\text{re} =\db{R} \cdot \vec{E}_t^\text{inc}$ with $ \vec{E}_t^\text{re}=\left( E_x^\text{re}  \; E_z^\text{re} \right)^T$. Then, the state vector $\vec{f}$ for $y\leq0$ can be expressed as: 
\begin{align} \label{E:field_left}
\vec{f}(y\leq 0)=
   \begin{pmatrix}
\left( \vec{1}_{2\times2}\e{ik_y y}+ \db{R}  \e{-ik_y y} \right) \cdot \vec{E}_t^\text{inc} \e{i\vec{k}_t\cdot\vec{r}}
     \\
   \frac{1}{\eta_0}\vec{J}\cdot \vec{Y}_\text{ext}   \cdot
  \left( \vec{1}_{2\times2}\e{ik_y y}-   \db{R} \e{-ik_y y} \right)\cdot \vec{E}_t^\text{inc} \e{i\vec{k}_t\cdot\vec{r}}
   \end{pmatrix}.
\end{align}
Similarly, for $y\geq d$, the  transmitted field can be expressed in terms of a transmission matrix as follows:
\begin{align}
\vec{f}(y\geq d)=
   \begin{pmatrix}
\db{T} \cdot \vec{E}_t^\text{inc} \e{i k_y (y-d)}\e{i\vec{k}_t\cdot\vec{r}}
     \\
    \frac{1}{\eta_0}\vec{J}\cdot \vec{Y}_\text{ext}   \cdot    \db{T} \cdot \vec{E}_t^\text{inc} \e{ik_y (y-d)}\e{i\vec{k}_t\cdot\vec{r}}
   \end{pmatrix}.
\end{align}
Inside the slab, the state vectors $\vec{f}(y)$  at $y=0$ and $y=d$ are linked by the relation \eqref{E:f_field_inside}. Then, enforcing the continuity of the tangential component of $\vec{E}$ and $\vec{H}$ at the interfaces $y=0$ and $y=d$, which is equivalent to enforce the continuity of the state vector $\vec{f}(y)$, we find that
\begin{align}
\e{-id\vec{M} }\cdot
   \begin{pmatrix}
\left( \vec{1}_{2\times2} + \db{R} \right) \cdot \vec{E}_t^\text{inc}  
     \\
    \frac{1}{\eta_0}\vec{J}\cdot \vec{Y}_\text{ext}  \cdot
  \left( \vec{1}_{2\times2} -   \db{R}  \right)\cdot \vec{E}_t^\text{inc}  
   \end{pmatrix} =
   \begin{pmatrix}
\db{T} \cdot \vec{E}_t^\text{inc} 
     \\
      \frac{1}{\eta_0}\vec{J}\cdot \vec{Y}_\text{ext}   \cdot   \db{T} \cdot \vec{E}_t^\text{inc} 
   \end{pmatrix}.
\end{align}
%
It is convenient to  write the transfer matrix as
\begin{align}
   \e{-id\vec{M}} =\begin{pmatrix}\db{A} &  \db{B}\\ \db{C}& \db{D}    \end{pmatrix} \label{E:ABCD_mat_MOSFET}
\end{align}
where $\db{A}, \,\db{B}, \, \db{C}$ and $\db{D}$ are $2\times2$ matrices. After some algebra, one can find the following explicit formulas for the reflection and transmission matrices:
\begin{align}
\db{R} &=  \left(\db{Q}_1+\db{Q}_2\right)^{-1} \cdot \left( \db{Q}_1-\db{Q}_2\right)\label{E:reflection_arbitrary_left}\\
 \db{T}&=2 \db{A} \cdot \left(\db{Q}_1+\db{Q}_2\right)^{-1} \cdot \db{Q}_1 + \frac{2}{\eta_0} \db{B} \cdot \vec{J}\cdot \vec{Y}_\text{ext} \cdot \left(\db{Q}_1+\db{Q}_2\right)^{-1} \cdot \db{Q}_2 \label{E:transmission_arbitrary_left} 
\end{align}
where
\begin{align}
 \db{Q}_1&= \frac{1}{\eta_0}\left( -  \db{D} \cdot \vec{J}\cdot \vec{Y}_\text{ext} +\frac{1}{\eta_0}\vec{J}\cdot \vec{Y}_\text{ext}   \cdot \db{B} \cdot \vec{J}\cdot \vec{Y}_\text{ext}  \right)\\
\db{Q}_2&= \db{C} - \frac{1}{\eta_0} \vec{J}\cdot \vec{Y}_\text{ext} \cdot \db{A}
\end{align}

For normal incidence ($k_x=k_z=0$) and a permittivity tensor as in \eqref{E:epsilon_example},  the transmission matrix [Eq. \eqref{E:transmission_arbitrary_left}] simplifies to Eq. \eqref{E:T_matrix} of the main text.

\newpage
\setcounter{section}{0}

\section*{References}
\bibliography{Biblio_CLEAN}

\end{document}